\documentclass[a4paper,fleqn,usenatbib]{mnras}
\usepackage[T1]{fontenc}
\usepackage{ae,aecompl}
\usepackage{graphicx}	
\usepackage{amsmath}	
\usepackage{amssymb}	
\usepackage{amsbsy}
\usepackage{float}
\usepackage{color}

\usepackage{mathrsfs,bm}
\newcommand{\bcdot}{\ensuremath{%
  \mathchoice%
   {\mskip\thinmuskip\lower0.2ex\hbox{\scalebox{1.5}{$\cdot$}}\mskip\thinmuskip}}%
   {\mskip\thinmuskip\lower0.2ex\hbox{\scalebox{1.5}{$\cdot$}}\mskip\thinmuskip}%
   {\lower0.3ex\hbox{\scalebox{1.2}{$\cdot$}}}%
   {\lower0.3ex\hbox{\scalebox{1.2}{$\cdot$}}}%
}

\def\del#1{{}}

\newcommand{\CR}{\rmn{cr}}
\newcommand{\eps}{\varepsilon}
\newcommand{\bnabla}{\ensuremath{\boldsymbol{\nabla}}}
\renewcommand{\d}{\rmn{d}}

\title[Anisotropic cosmic ray transport on an unstructured mesh]{Semi-implicit anisotropic cosmic ray transport on an unstructured moving mesh}

\author[R.~Pakmor et al.]  {R\"udiger~Pakmor$^1$\thanks{E-mail: ruediger.pakmor@h-its.org},
  Christoph~Pfrommer$^{1}$, Christine~M.~Simpson$^{1}$, Rahul~Kannan$^{2}$, \newauthor Volker~Springel$^{1,3}$\vspace*{0.2cm}\\
  $^1$Heidelberger Institut f\"{u}r Theoretische Studien,
  Schloss-Wolfsbrunnenweg 35, 69118 Heidelberg, Germany\\
  $^2$Department of Physics, Kavli Institute for Astrophysics $\&$ Space Research,
  Massachusetts Institute of Technology, \\ Cambridge 02139, MA, USA \\
  $^3$Zentrum f\"ur Astronomie der Universit\"at Heidelberg,
  Astronomisches Recheninstitut, M\"{o}nchhofstr. 12-14, 69120
  Heidelberg, Germany \\
}
  
\pubyear{2016}

\begin{document}

\label{firstpage}
\pagerange{\pageref{firstpage}--\pageref{lastpage}}

\maketitle

\begin{abstract}
  In the interstellar medium of galaxies and the intracluster gas of
  galaxy clusters, the charged particles making up cosmic rays are
  moving almost exclusively along (but not across) magnetic field
  lines.  The resulting anisotropic transport of cosmic rays in the
  form of diffusion or streaming not only affects the gas dynamics but
  also rearranges the magnetic fields themselves. The coupled dynamics
  of magnetic fields and cosmic rays can thus impact the formation and
  evolution of galaxies and the thermal evolution of galaxy clusters
  in critical ways. Numerically studying these effects requires
  solvers for anisotropic diffusion that are accurate, efficient, and
  robust, requirements that have proven difficult to satisfy in
  practice. Here, we present an anisotropic diffusion solver on an
  unstructured moving mesh that is conservative, does not violate the
  entropy condition, allows for semi-implicit time integration with
  individual timesteps, and only requires solving a single linear
  system of equations per timestep. We apply our new scheme to a large
  number of test problems and show that it works as well or better
  than previous implementations. Finally, we demonstrate for a
  numerically demanding simulation of the formation of an isolated
  disk galaxy that our local time-stepping scheme reproduces the
  results obtained with global time-stepping at a fraction of the
  computational cost.
\end{abstract}


\section{Introduction}

Energetic particles carry a significant fraction of the free energy in
the Universe. Their acceleration, their interactions, and their
eventual fate are integral parts of the evolution of astrophysical
systems such as galaxies, galaxy clusters, and the intergalactic
medium that fills the space in between.  Particle accelerators for
these charged, relativistic particles, called cosmic rays (CRs), are
strong shocks, forming for example when a star explodes as a
supernova, when a black-hole system ejects matter at relativistic
velocities, or when galaxy clusters collide \citep{Blandford1987}. In
the interstellar medium (ISM), shocks at supernova remnants are able
to accelerate a tiny fraction of $\sim10^{-9}$ particles to
relativistic energies so that the resulting population of CRs collects
on average an equal share of the total energy stored in thermal
particles, turbulence, and magnetic fields \citep{Boulares1990}.

The processes that determine the energy and spatial distribution of
CRs are different from those that shape collisional gases, because
they rely almost entirely on electric and magnetic fields: the CR
particle mean free path due to collisional processes (which are
dominated by hadronic interactions with the ambient gas at
relativistic CR energies) is
$\lambda_{\rmn{had}}\sim 10\,n_{\rmn{cm}^{-3}}^{-1}\,\rmn{Mpc}$ and thus
much larger than any relevant scale in galaxies or clusters
\citep{Pfrommer2004}.  The potential wells of galaxies
($1$~to~$100$~eV) and galaxy clusters ($1$~to~$10$~keV) would be too
shallow to keep relativistic particles bound if they could move freely
along straight lines. The escape time of CRs from the ISM of the
Galaxy is however quite long, of order $3\times10^7$~yr. Also, the
observed spatial isotropy of CRs \citep{Schlickeiser2002, Kulsrud2005}
suggests the existence of a very efficient scattering mechanism that
isotropises their distribution.

These scattering processes support a kind of self-organisation that,
through interactions between particles and electromagnetic fields,
arranges the particles and the available energy into three components:
a cool or warm gas phase that carries the bulk of the mass, energetic
particles with a wide range of energies, and a turbulent
electromagnetic field that links the two. CRs streaming along the
magnetic field excite resonant Alfv\'en waves through the `streaming
instability' \citep{Lerche1967, Kulsrud1969, Skilling1971}. As a
result, the CRs experience rapid pitch-angle scatterings with this
wave field, which causes an isotropization of their distribution in
the frame of the Alfv\'en waves, thereby limiting the CR bulk speed
to the Alfv\'en velocity.  In addition, CRs can diffuse relative to
this wave frame, as well as across field lines by scattering off
magneto-hydrodynamic (MHD) turbulence \citep{Wiener2013}. 

However, interactions with Alfv\'en modes of the cascade are extremely
inefficient due to the increasing anisotropy at small scales, with
power concentrated in modes with wave-vectors transverse to the
magnetic field, while CRs efficiently scatter off the parallel
component \citep{Chandran2000, Yan2002}. Fast magnetosonic modes could
potentially scatter CRs more efficiently \citep{Yan2004,
  Brunetti2007}, but this process is thought to be a small effect
\citep{Desiati2014} so that CRs are transported almost entirely along
magnetic field lines, giving rise to strongly anisotropic CR transport.

What is the observational evidence for anisotropic CR transport?
Observations of polarised radio emission from edge-on galaxies show
poloidal field lines at the interface of the galactic disk and halo
\citep[e.g.,][]{Tuellmann2000}, which corresponds to the location from
where galactic winds are launched. This argues for a dynamical
mechanism that is able to reorient the preferentially toroidal field
in galactic disks by means of anisotropic CR transport along the
magnetic field. If CR energy is injected into the ISM by diffusive
shock acceleration at supernova remnants, the light CR components
finds itself underneath the mass-carrying thermal phase of the ISM. As
a consequence, CRs attempt to buoyantly rise from the disk and drive a
Parker instability \citep{Rodrigues2016}. As they are dragging the
magnetic field with them, this eventually results in a poloidal
(`open') field configuration.

In recent years, different groups have started to incorporate CR transport into
simulations of galaxy formation and evolution and simulations of the ISM. Simulations of the
formation of isolated disk galaxies have demonstrated the presence of strong
CR-driven outflows by following isotropic \citep{Uhlig2012} or anisotropic
\citep{Ruszkowski2016} CR streaming as the active transport mechanism for
CRs. It was also demonstrated that isotropic CR diffusion can drive galactic
winds in isolated Milky Way-sized disk galaxies \citep{Booth2013,
  Salem2014}. Moreover, this result still holds for a simulated galaxy that
forms in a cosmological environment \citep{Salem2014b}. On smaller scales,
simulations of the ISM that account for anisotropic CR diffusion demonstrate the
launch of strong winds from the disk \citep{Hanasz2013, Girichidis2016}.

All these simulations employ an explicit time integration scheme. This
severely limits the achievable resolution since the timestep
constraint of explicit schemes for CR diffusion scales quadratically
with linear dimensions of a resolution element. Thus, it is desirable
to develop semi-implicit or implicit time-integration schemes in order
to avoid a very restrictive timestep limitation of the diffusion solver.

Moreover, \citet{Sharma2007} recently analysed the numerical
properties of discretisation schemes for anisotropic heat conduction
(which exhibit the same mathematical properties as anisotropic
diffusion) and demonstrated that they generally violate the entropy
condition $\Delta S \geq 0$, i.e., heat can be transported from cold
to warm regions (or equivalently CRs are transported from regions of
low to high abundance), unless this behaviour is not explicitly
prevented by an ingenious design of the numerical
scheme. Unfortunately, anisotropic diffusion schemes that do not
violate the entropy condition introduce non-linearities that
significantly increase the numerical complexity when one tries to 
combine them with an efficient implicit time integrator, although
efficient {\em semi-}implicit schemes are still possible
\citep{Sharma2011,Kannan2016,Hopkins2016DiffAniso}.

Of particular importance for multi-scale problems such as galaxy
evolution is another issue that arises for implicit time integration
schemes from the coupling to the dynamics of the magnetic field. To
take the dynamics of the magnetic field accurately into account in the
diffusion solver, the diffusion problem for every cell (or particle)
has to be integrated on a timestep smaller or equal to its MHD
timestep. Thus, integrating the diffusion solver on global timesteps
requires a reduction of the global timestep to the smallest MHD
timestep in the simulation, rendering large, multi-scale problems to
become almost impossible. To avoid this impasse, we must develop a
method that allows for individual timesteps, i.e., it needs to
tolerate different diffusion timesteps for each cell chosen on local
conditions, while at the same time guaranteeing stability for
timesteps that are much larger than those required by the stability
criterion of {\em explicit} time integration of the diffusion
problem. Such a scheme has been proposed by \citet{Dubois2016},
however, their algorithm does not ensure energy conservation and in
general violates the entropy condition.

Here, we discuss a novel anisotropic diffusion scheme implemented in
the moving-mesh code \textsc{arepo} \citep{Arepo}. It is based on the
flux limiting scheme by \citet{Sharma2007} and the semi-implicit
time-integration scheme presented by \citet{Sharma2011}, with some
refinements. In particular, we generalise the scheme to unstructured
meshes and extend it to allow for semi-implicit time integration with
individual timesteps while still ensuring energy conservation and
preserving the entropy condition. The present paper concentrates on
the technical treatment of the CR diffusion problem, while a
companion paper \citep{Pfrommer2016} describes the foundations of
the CR physics implemented in the code, focusing in particular on the
source and sink processes. In two further companion papers, we apply
our new methods to wind formation in disk galaxies \citep{Pakmor2016b}
and to the effects of individual supernova explosions in the ISM
\citep{Simpson2016}.

This paper is structured as follows.  We first describe our algorithm
and implementation in Section~\ref{sec:implementation}. We then apply
it to several common test problems in Section~\ref{sec:tests} and show
that the implicit individual time-stepping delivers exquisite results
for the problem of the formation of an isolated disk galaxy in
Section~\ref{sec:galaxy}. Finally, we summarise our results in
Section~\ref{sec:summary}.
   
\section{Implementation}

\label{sec:implementation}

The part of the evolution equation for the CR energy density
$\eps_\CR$ that describes anisotropic diffusion is given by
\begin{equation}
\frac{\partial \eps_\CR}{\partial t} - \bnabla \bcdot \left[ \kappa_\eps \mathbfit{b} \left( \mathbfit{b} \bcdot \bnabla \eps_\CR \right) \right] = 0,
\end{equation} 
where $\eps_\CR$ is the CR energy density, $\kappa_\eps$ is the
kinetic energy-weighted diffusion coefficient, and
$\mathbfit{b} = \mathbfit{B}/\left| \mathbfit{B} \right|$ is the
direction of the magnetic field. The full evolution equation for the
CR energy density includes terms to describe the acceleration, the
advection, and the cooling of CRs \citep{Pfrommer2016}\footnote{Unlike \citet{Pfrommer2016} Sec.~3.3, here we use the collisional heating rate due to
Coulomb interactions only, where $\Gamma_\mathrm{th} = - \Lambda_{\mathrm{Coul}} = \tilde\lambda_\mathrm{th} n_\mathrm{e} \varepsilon_\mathrm{cr}$ and
$\tilde \lambda_\mathrm{th} =2.78 \times 10^{-16}~\mathrm{cm}^3~\mathrm{s}^{-1}$.}. Here, we
concentrate only on the diffusion term; the other terms are
effectively treated in an operator split fashion by the code.

After integrating the diffusion term over the volume $V_i$ of cell
$i$, applying Gauss' theorem to the divergence term, and dividing by
the cell volume, we obtain
\begin{equation}
\frac{\partial \eps_{\CR,i}}{\partial t} - \frac{1}{V_i} \int_{\partial V_i} \kappa_\eps \left( \mathbfit{b} \bcdot \bnabla \eps_\CR \right)\mathbfit{b} \bcdot \d \mathbfit{A}_i = 0.
\end{equation}
Here, $\eps_{\CR,i}$ is the CR energy density in cell $i$ and $\d \mathbfit{A}_i$ points along the normal vector of the surface.

Discretization of the surface integral leads to
\begin{equation}
\frac{\partial \eps_{\CR,i}}{\partial t} -  \frac{1}{V_i} \sum_{j} \kappa_{ij} \left( \mathbfit{b}_{ij}\bcdot \bnabla \eps_{\CR,ij} \right) \mathbfit{b}_{ij} \bcdot \mathbfit{n}_{ij}A_{ij} = 0.
\label{eq:crdiff}
\end{equation}
This sum runs over all interfaces $j$ of cell $i$ and depends on the
diffusion coefficient $\kappa_{ij}$, the direction of the magnetic
field $\mathbfit{b}_{ij}$, the gradient of the CR energy density
$\bnabla \eps_{\CR,ij}$ \textit{at the interface}, the normal vector
$\mathbfit{n}_{ij}$ of the interface and its area $A_{ij}$. Note that
for the simpler case of isotropic diffusion the equation becomes
\begin{equation}
\frac{\partial \eps_{\CR,i}}{\partial t} - \frac{1}{V_i}  \sum_{j} \kappa_{ij} \left(\bnabla \eps_{\CR,ij} \bcdot\mathbfit{n}_{ij}\right) A_{ij} = 0.
\end{equation}

To evaluate the sum in Eq.~(\ref{eq:crdiff}) we need an estimate of
the full gradient of the CR energy density at every
interface. Unfortunately, a na\"{\i}ve estimate of this gradient can
lead to unphysical solutions, as one can easily construct situations
where the energy can flow from a cell with lower energy density to
a cell with higher energy density. One way to avoid this unphysical
solution is to limit the gradient appropriately, as described by
\citet{Sharma2007} for a Cartesian mesh. We follow a similar strategy 
for tackling this problem, but one that is generalized for
an unstructured, Voronoi mesh.

\begin{figure*}
  \centering
  \begin{minipage}[b]{0.45\textwidth}
    \centering
    \includegraphics[width=\textwidth]{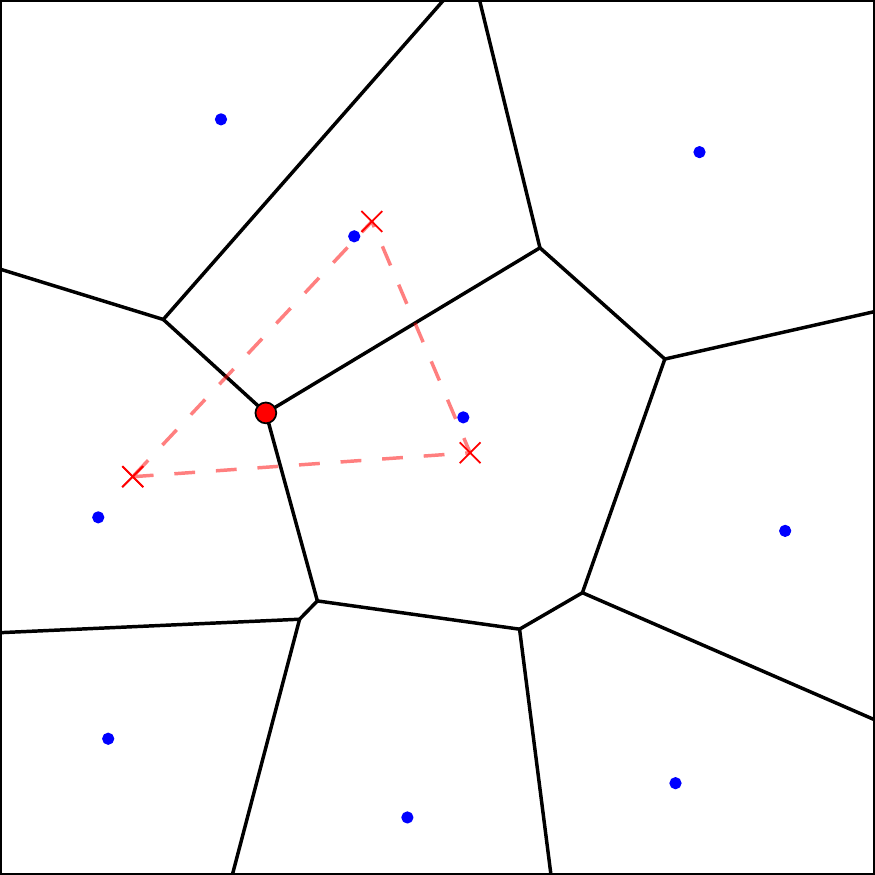}
  \end{minipage}
  \hfill
  \begin{minipage}[b]{0.45\textwidth}
    \centering
    \includegraphics[width=\textwidth]{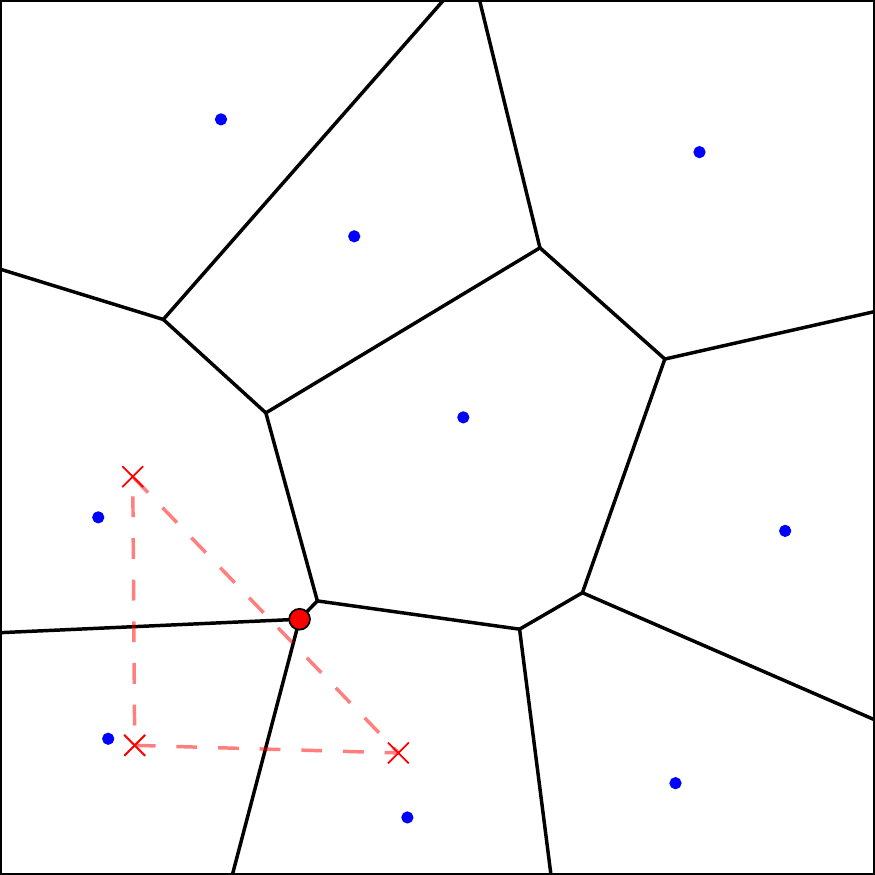}
  \end{minipage}
  \caption{Visualisation of the gradient estimate at a corner of the
    Voronoi mesh in 2D. Blue circles show the positions of the
    mesh-generating points, black lines show the edges of the Voronoi
    cells. The red circle in each panel shows a selected corner whose
    gradient estimate is based on the values at the centers of mass of
    the three adjacent cells marked by red crosses. The left panel
    shows a well-behaved configuration, i.e.~the corner lies within
    the triangle spanned by the three centers of mass. In contrast,
    the right panel shows an ill-behaved configuration for which the
    desired interpolation turns into an extrapolation because the
    corner lies outside the triangle.}
  \label{fig:corners}
\end{figure*}

\subsection{Estimating the gradient at a corner}
\label{sec:cornergrads}

Similar to \cite{Guenter2005} and \cite{Sharma2007}, we build the gradient estimate at the
center of an interface based on gradient estimates of its corners. Thus,
we start with gradient estimates at the corners of the Voronoi
mesh. In 2D, every corner of a Voronoi mesh connects three cells. The
mesh-generating points of these cells create a triangle whose
circumcircle center is the position of the corner. To estimate the
gradient of a quantity at the corner, we use its values at the centers
of mass of the three adjacent cells as sketched in Fig.~\ref{fig:corners}
and perform a least squares fit on the value of the quantity and its
gradient at the corner.

In 3D, the triangle becomes a tetrahedron and every corner has four
adjacent cells. The residuals of the fit are given by
\begin{equation}
r_i = \phi \left( \mathbfit{c} \right) + \left( \bnabla \phi \right) \left( \mathbfit{s}_i - \mathbfit{c} \right) - \phi \left( \mathbfit{s}_i \right),
\end{equation}
where $\phi \left( \mathbfit{c} \right)$ is the unknown value at the position of the corner $\mathbfit{c}$, $ \bnabla \phi$ is the gradient at the corner, and $\phi \left( \mathbfit{s}_i \right)$ is the value at the center $\mathbfit{s}_i$ of cell $i$. We can rewrite this in the form of a matrix equation
\begin{equation}
\mathbfit{r} = \mathbfss{X} \mathbfit{q} - \mathbfit{Y}.
\end{equation}
Here, $\mathbfit{r}$, $\mathbfit{q}$, and $\mathbfit{Y}$ are N-vectors with $N=3$ ($N=4$), and $\mathbfss{X}$ is a $N\times N$ square matrix in 2D (3D). They are defined as
\begin{eqnarray*}
(\mathbfit{q})_0 &=& \phi \left( \mathbfit{c} \right) \\
(\mathbfit{q})_{1..N-1} &=& \left( \bnabla \phi \right)_{0..N-2} \\
(\mathbfit{Y})_i &=& \phi \left( \mathbfit{s}_i \right) \\
(\mathbfss{X})_{i,0} &=& 1 \\
(\mathbfss{X})_{i,1..N-2} &=& \mathbfit{s}_{i,0..N-2} -  \mathbfit{c}_{0..N-2}.
\end{eqnarray*}

To minimize the residual we solve the normal equations and find
\begin{equation}
\left( \mathbfss{X}^T\ \mathbfss{X} \right) \mathbfit{q} = \mathbfss{X}^T\ \mathbfit{Y}.
\end{equation}
Since $\mathbfss{X}$ is a square matrix, there is a unique solution for $\mathbfit{q}$ with zero residual. To solve for $\mathbfit{q}$ we multiply with $\left( \mathbfss{X}^T\ \mathbfss{X} \right)^{-1}$ from the left and obtain
\begin{equation}
$\mathbfit{q}$ = \mathbfss{M} \ \mathbfit{Y},
\label{eq:gradcorner}
\end{equation}
with
$\mathbfss{M} = \left( \mathbfss{X}^T\ \mathbfss{X} \right)^{-1} \
\mathbfss{X}^T$.
Note that $\mathbfss{M}$ only depends on the geometry of the mesh and
$\mathbfit{Y}$ only contains values at the centers of cells. Thus, we
only have to compute $\mathbfss{M}$ once in every timestep because the
geometry of the mesh does not change and can obtain the value and
gradient of any quantity at a corner from a simple matrix-vector
multiplication. Moreover, Eq.~(\ref{eq:gradcorner}) describes the
linear dependence of the gradient at the corner on the values in the
adjacent cells. This is needed later on for the implicit time
integration.

The value of a quantity at the corner only depends on the entries in
the first row of $\mathbfss{M}$. If the corner lies inside the
triangle (tetrahedron) spanned by the centers of mass of the adjacent
cells, all values in the first row of $\mathbfss{M}$ are positive and
the value of a quantity at the corner is interpolated from the values
at the centers of the adjacent cells. However, as demonstrated in
Fig.~\ref{fig:corners}, for general mesh geometries it is possible
that the corner lies outside the triangle (tetrahedron). In this case,
at least one of the entries in the first row of $\mathbfss{M}$ is
negative and the interpolation becomes an extrapolation, which can in
general be unstable. To deal with this we mark any corner as
problematic which has an entry in the first row of
$\mathbfss{M}_{0,j} < -0.01$ and treat these corners in a special way,
as described in Section~\ref{sec:flux}. One might consider
aggressively setting this threshold to zero, thus not allowing for any
extrapolation at the corners at all. However this turns out to
increase the diffusivity of the scheme perpendicular to the magnetic
field vector and does not seem to be necessary for any of our test
problems.

\subsection{Calculating the direction of the magnetic field at the interface}

\begin{figure}
  \centering
  \includegraphics[width=\linewidth]{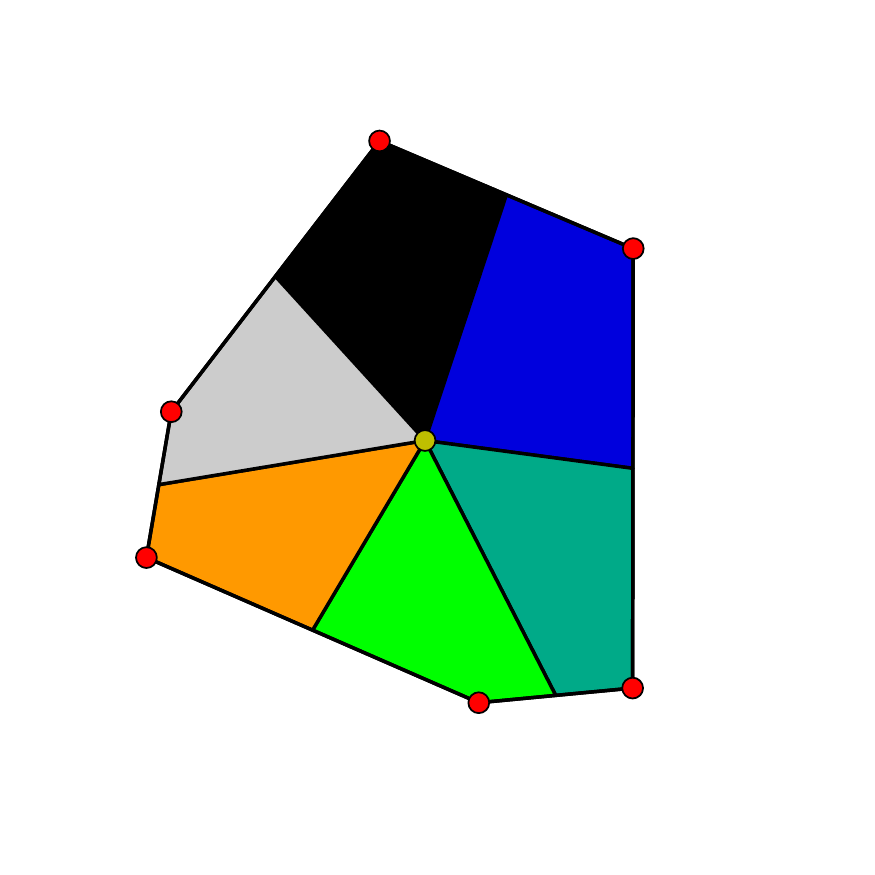}
  \caption{Sketch of the geometric weights with which the different
    corners contribute to the value of the magnetic field and other
    quantities at the center of the interface in 3D. The red circles
    are the corners of the interface, the yellow circle shows the
    position of the center of the interface, and the size of the
    coloured areas is proportional to the weight that is associated
    with the corners.}
  \label{fig:face}
\end{figure}

The least squares fit estimate of the gradient at the corners also
allows us to compute the value of a quantity at the position of a
corner. To then obtain a single value for the interface, we perform a
weighted average of these values at the corners,
\begin{equation}
\phi_\mathrm{face} = \sum_i w_i \phi \left( \mathbfit{c}_i \right).
\label{eq:faceaverage}
\end{equation}
The weights are set to
\begin{equation}
w_i = \frac{A_i}{ A_\mathrm{face} },
\end{equation}
where $A_\mathrm{face}$ is the area of a face, and $A_i$ the area of a
tetrahedron defined by the position of the corner, the center of the
face, and the two midpoints of the two adjacent edges of the corners
that are part of the face (see Fig.~\ref{fig:face}). In 2D, this
simplifies to $w_i = 0.5$ for both corners of an interface.

\subsection{Computing the flux}
\label{sec:flux}

To compute the flux over an interface as given by
Eq.~(\ref{eq:crdiff}) we rotate the coordinate system so that one axis
points along the normal vector of the interface, one axis points along
the magnetic field direction in the interface, and the third axis is
perpendicular to the first two, while lying in the interface. We then
split the gradient of the CR energy density at the interface into two
components: one along the normal direction of the interface and the
other in the plane of the interface. In 3D, the latter in principle is
a two-dimensional vector. However, we rotate the tangential component
in the interface such that one component points along the magnetic
field in the interface, and ignore the remaining component that is in
the interface but perpendicular to the magnetic field vector and
therefore always exactly zero. Thus we can write the split in 2D and
in 3D as
\begin{equation}
\bnabla \eps_\CR = \bnabla \eps_{\CR\rmn{,n}} + \bnabla \eps_{\CR\rmn{,t}}.
\end{equation}
We then compute two separate fluxes from them, which add up to the total flux.

We can prevent unphysical fluxes from a cell with lower CR energy
density to a cell with higher CR energy density by limiting the
gradient on the interface \citep{Sharma2007}. We use a generalized
version of the van-Leer limiter \citep{vanLeer1984} to compute the
limited gradient estimate $\bnabla \eps_{\CR\rmn{,t}}$ along the
magnetic field direction in an interface from the unlimited gradients
in the same direction $\bnabla \eps^i_{\CR\rmn{,t}}$ at corner $i$. If
all values of $\bnabla \eps^i_{\CR\rmn{,t}}$ have the same sign, we
calculate $\bnabla \eps_{\CR\rmn{,t}}$ by
\begin{equation}
\bnabla \eps_{\CR\rmn{,t}} = \frac{ N }{ \sum_i \left(\bnabla \eps^i_{\CR\rmn{,t}}\right)^{-1} }.
\end{equation}
\noindent If they do not all have the same sign, we set
$\bnabla \eps_{\CR\rmn{,t}} = 0$. 

The gradient estimate includes all
corners of the interface except for the ones we previously marked as
problematic because their gradient estimate became an extrapolation,
as described in detail in Section~\ref{sec:cornergrads}. Knowing
$\bnabla \eps_{\CR\rmn{,t}}$ we can directly compute its associated
flux. In the rare case that all corners of an interface are flagged,
we use the stricter minmod limiter instead and include all corners.

We implement two different ways to compute the gradient of the CR
energy density along the normal direction of the interface. We
refer to them as \textit{simple} normal gradients and \textit{full} normal
gradients. The \textit{simple} normal gradient estimate only takes the
values at centers of mass of the two adjacent cells of an interface into
account. From those, we compute a one-dimensional gradient along the
connecting line between the two centers (L,R) and project it onto the
normal vector of the interface. The gradient estimate is then given by
\begin{equation}
 \bnabla \eps_{\CR\rmn{,n}} =  \frac{\eps_\mathrm{CR,L} - \eps_\mathrm{CR,R}}{ \left| \mathbfit{c}_L - \mathbfit{c}_R \right| } 
 \left(\frac{ \mathbfit{c}_L - \mathbfit{c}_R }{ \left| \mathbfit{c}_L - \mathbfit{c}_R \right| }\right) \bcdot\mathbfit{n}_\mathrm{face}.
\end{equation}

If the centers of mass of the two cells are at the same position as
their mesh-generating points, this estimate is quite accurate because
the normal vector of the interface is parallel to the connecting line
between the two mesh-generating points. However, in general there is a
small deviation between the mesh-generating point and the center of
mass of a cell. In this case, the simple normal gradient estimate
slightly underestimates the true normal gradient at the interface.

For the \textit{full} normal gradient estimate on an interface we
again make use of the gradient estimates at its corners. We use the
same averaging procedure for the normal components of the corner
gradients that we use for the magnetic field values at the interface
(see Eq.~\ref{eq:faceaverage}). However, here we ignore the
contributions of all corners that have been flagged as problematic and
replace their contribution to the total sum with the simple gradient
estimate.

We do not need to limit the \textit{simple} normal gradients, because
they ensure the entropy condition by construction. When we use
\textit{full} normal gradients, we check for every interface if the
direction of the normal gradient given the current CR energy densities
of the surrounding cells violates the entropy condition. If that is
the case, we use the \textit{simple} normal gradient estimate instead
for this interface.

Note that both estimates of the normal component of the gradient are
linear in their dependence on the CR energy densities of the
cells. However, for the full gradient estimate the normal component of
the gradient for an interface usually depends on the values of a dozen
cells or more. In comparison, the simple gradient estimate for an
interface only depends on two cells.

\subsection{Time integration}

We implement two different time integration schemes for anisotropic
diffusion, a purely explicit scheme and a semi-implicit scheme. For
the explicit time integration, the update is done in two half-steps
\begin{equation}
\begin{aligned}
  \eps_{\CR,i}^{n+1/2} = \eps_{\CR,i}^{n} &+ \frac{\Delta t}{2 V_i} \left[\sum_{j} \kappa_{ij}
  \left( \mathbfit{b}_{ij} \bcdot\bnabla \eps_{\mathrm{CR,t,}ij}^{n} \right) \mathbfit{b}_{ij} \bcdot\mathbfit{n}_{ij} A_{ij}\right. \\
  & + \left.\sum_{j} \kappa_{ij} \left( \mathbfit{b}_{ij} \bcdot\bnabla \eps_{\mathrm{CR,n,}ij}^{n} \right) \mathbfit{b}_{ij} \bcdot\mathbfit{n}_{ij} A_{ij}\right], \\
    \eps_{\CR,i}^{n+1} = \eps_{\CR,i}^{n+1/2} &+ \frac{\Delta t}{2 V_i} \left[\sum_{j} \kappa_{ij}
      \left( \mathbfit{b}_{ij} \bcdot\bnabla \eps_{\mathrm{CR,t,}ij}^{n+1/2} \right) \mathbfit{b}_{ij} \bcdot\mathbfit{n}_{ij} A_{ij}\right. \\
      &+ \left. \sum_{j} \kappa_{ij} \left( \mathbfit{b}_{ij} \bcdot\bnabla \eps_{\mathrm{CR,n,}ij}^{n+1/2} \right)  \mathbfit{b}_{ij} \bcdot\mathbfit{n}_{ij} A_{ij}\right].
\end{aligned}
\end{equation}
The two steps are applied at the beginning and at the end of each
timestep, respectively.  $\eps_{\CR,i}^{n}$, $\eps_{\CR,i}^{n+1/2}$,
and $\eps_{\CR,i}^{n+1}$ denote the CR energy density in cell $i$ at
the beginning of the timestep, in the middle of the timestep, and at
the end of the timestep, respectively.

Moreover, to improve the order of accuracy of the time integration, we
first integrate the diffusion term for a half-timestep at the
beginning of the timestep, before the first part of the gravity
timestep. Later, we integrate it for another half-timestep at the end
of the timestep, after the second part of the gravity
timestep. Unfortunately, a timestep constraint
\begin{equation}
\Delta t_\mathrm{diffusion} < \frac{\Delta x^2}{\kappa_\eps},
\end{equation}
where $\Delta x$ is the diameter of a cell and $\kappa_\eps$ the
diffusion coefficient in a cell, is required for stability, which
severely limits the applicability of the explicit time integration
scheme. A common solution to guarantee stability even for large
timesteps is to switch to an implicit time integration
scheme. However, the limiting procedure for the gradients in the
interface makes the system nonlinear and very expensive to solve in a
fully implicit scheme.

\subsection{Semi-implicit time integration}
Fortunately, it is possible to formulate a semi-implicit scheme that
is almost as stable as the fully implicit scheme, but only requires 
solving one linear implicit problem per timestep \citep{Sharma2011}. To
achieve this, we split the time integration into two parts. We first
advance the CR energy of all cells with the fluxes associated with the
components of the CR energy density gradients in the interfaces using
an explicit forward Euler time integrator, because the nonlinearity of
the tangential gradient estimate $\eps_{\mathrm{CR,t},ij}^{n}$
prohibits an efficient implicit solution, as
\begin{equation}
\begin{aligned}
\eps_{\CR,i}^{\tilde{n}} = \eps_{\CR,i}^{n} +  \frac{\Delta t}{V_i}  \sum_{j} \kappa_{ij} \left( \mathbfit{b}_{ij} \bcdot\bnabla \eps_{\mathrm{CR,t},ij}^{n} \right) \mathbfit{b}_{ij} \bcdot\mathbfit{n}_{ij} A_{ij}.
\end{aligned}
\end{equation}

In a second step, we advance the CR energy of all cells with the
fluxes associated with the normal component of the CR energy density
gradients at the interfaces that are linear.  This is done using an
implicit backward Euler scheme:
\begin{equation}
\begin{aligned}
\eps_{\CR,i}^{n+1} = \eps_{\CR,i}^{\tilde{n}} +  \frac{\Delta t}{V_i}   \sum_{j} \kappa_{ij} \left( \mathbfit{b}_{ij} \bcdot\bnabla \eps_{\mathrm{CR,n},ij}^{n+1} \right) \mathbfit{b}_{ij} \bcdot\mathbfit{n}_{ij} A_{ij}.
\end{aligned}
\end{equation}

Since $\bnabla \eps_{\mathrm{CR,n},ij}^{n+1}$ depends only linearly on
the CR energy densities of a set of cells, this constitutes a system
of coupled linear equations that can be solved with a matrix
solver. We solve the linear system in a two-step procedure using
solvers from the \textsc{hypre} library \citep{HYPRE}. We first try to
solve the system with the iterative, flexible GMRES solver
\citep{Saad1986} until we either reach a residual smaller than
$10^{-8}$ or exceed $200$ iterations. In the latter case, we add an
algebraic multigrid preconditioner \citep{Henson2002} to GMRES and
iterate until the residual is smaller than $10^{-8}$. This two-step
procedure is often significantly faster than using the multigrid
preconditioner directly, as the setup phase of the multigrid solver
can take more time than $200$ iterations of the pure GMRES
solver. Note, however, that for some problems it may be more efficient
to always run with the multigrid preconditioner.

Similarly to the findings of \citet{Sharma2011}, the scheme seems to
be stable for very large timesteps, but the result eventually becomes
incorrect and the entropy condition can be violated. Moreover, for
large diffusion coefficients and large timesteps, significant errors
may occur at the timestep boundaries. In practice, this does not
happen for cosmic ray diffusion as the cosmic ray diffusion
coefficient is somewhat limited, but it may become an issue for
certain problems of heat diffusion. If the scheme is applied in such
cases, it may require additional constraints on the diffusion
timestep.

\begin{figure}
  \centering
  \includegraphics[width=\linewidth]{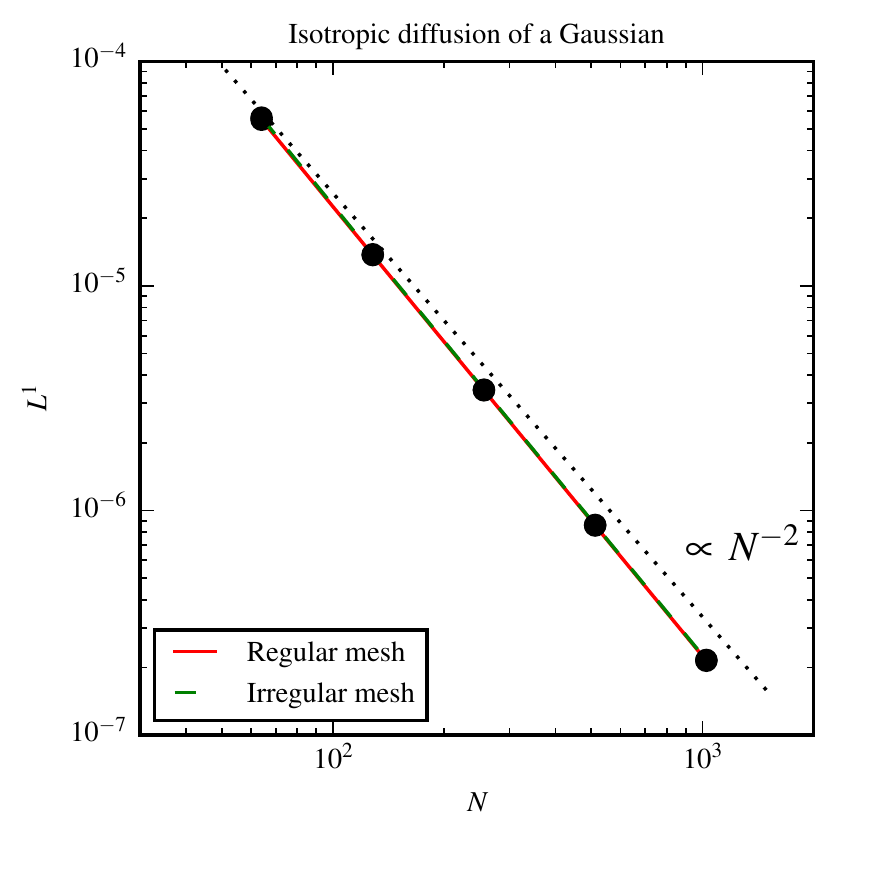}
  \caption{$L^1$ norm for isotropic diffusion of a 2D Gaussian
    profile. The red and green lines show the norm for a regular and
    an irregular mesh, respectively.}
  \label{fig:L1IsoGauss}
\end{figure}

\begin{figure}
  \centering
  \includegraphics[width=\linewidth]{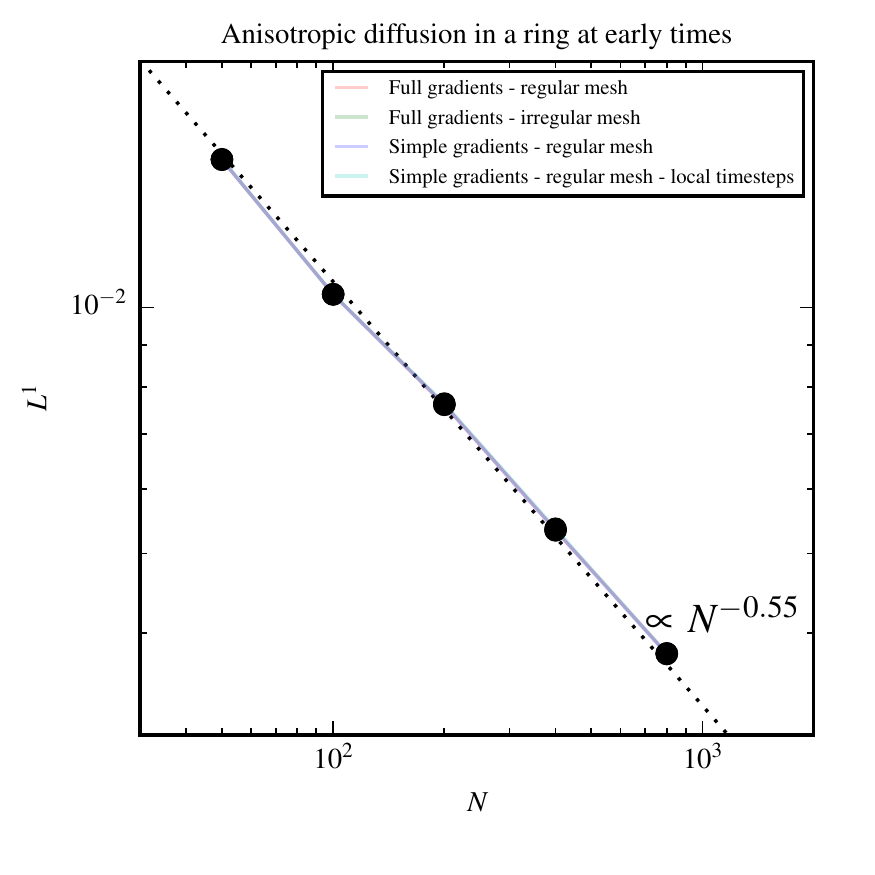}
  \includegraphics[width=\linewidth]{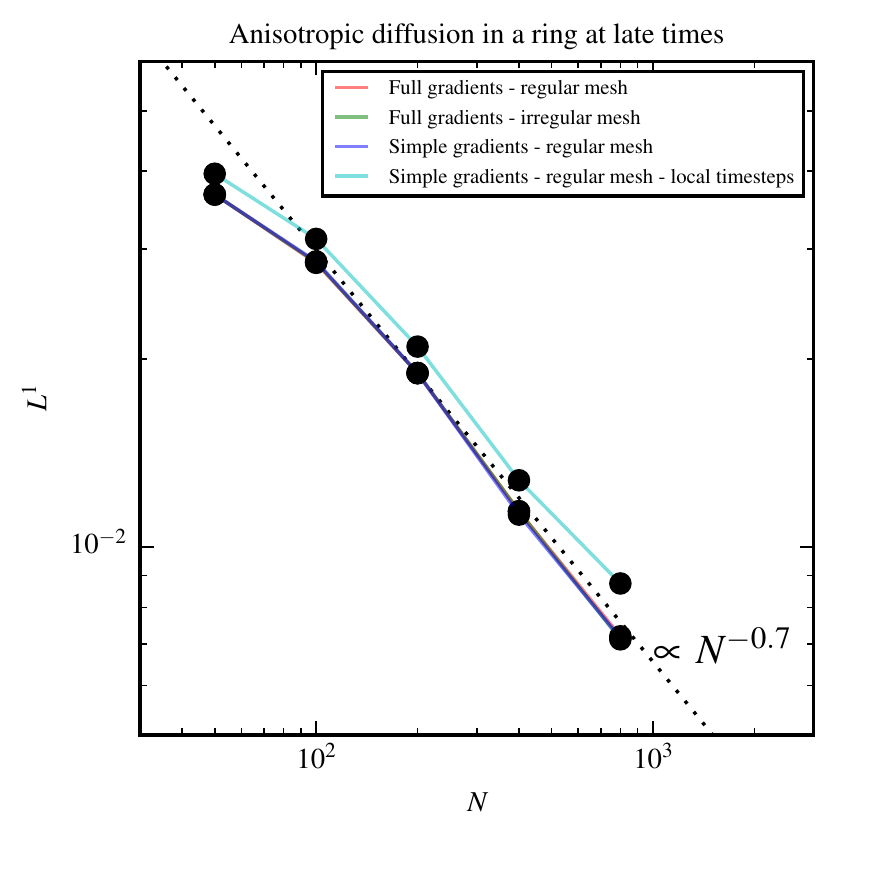}
  \caption{$L^1$ norm for anisotropic diffusion of an initial wedge in
    a circular magnetic field in 2D at times $t=10$ (top panel) and $t=200$
    (bottom panel). The solid lines show runs that use different
    gradient estimates, different meshes, or different time
    integration schemes, as labelled.}
  \label{fig:L1AnIsoRing}
\end{figure}

For isotropic diffusion we use a Crank-Nicolson scheme in a similar
fashion, i.e.~we split the flux into two half-timestep fluxes. We then
integrate the first flux explicitly and the second flux implicitly:
\begin{equation}
\begin{aligned}
\eps_{\CR,i}^{\tilde{n}} &= \eps_{\CR,i}^{n} +  \frac{\Delta t}{2 V_i}  \sum_{j} \kappa_{ij} \left(\bnabla \eps_{\mathrm{CR,n},ij}^{n} \bcdot \mathbfit{n}_{ij} \right) A_{ij},\\
\eps_{\CR,i}^{n+1} &=\eps_{\CR,i}^{\tilde{n}} +  \frac{\Delta t}{2 V_i}  \sum_{j} \kappa_{ij} \left(\bnabla \eps_{\mathrm{CR,n},ij}^{n+1} \bcdot \mathbfit{n}_{ij}\right) A_{ij}.
\end{aligned}
\end{equation}

\subsection{Semi-implicit local timestepping}

Even though the semi-implicit time integration scheme is stable for
large timesteps, for anisotropic diffusion timesteps larger than the
smallest MHD timestep the diffusion does not take into account the
full dynamics of the magnetic field that generally changes on a MHD
timestep. However, reducing the global diffusion timestep to the
smallest MHD timestep will make problems with a deep hydrodynamical
timestep hierarchy (e.g., simulations of galaxy formation)
unfeasible. Therefore, we modify our semi-implicit time integration
scheme such that in one timestep it only solves the diffusion problem
for the active cells, as determined by the MHD timestep and for a layer
of boundary cells. In this way, we are able to combine the stability of
the semi-implicit time integrator with the efficiency of an individual
timestepping approach.

Our approach follows the time integration scheme for individual
timesteps for the hydrodynamics solver implemented in \textsc{arepo}
\citep{Arepo,Pakmor2016}. For every timestep we adopt the following
procedure:
\begin{enumerate}
\item We create a list of all active interfaces, i.e.~interfaces with
  at least one adjacent active cell.
\item We then collect all cells that share at least one corner with an
  active interface. This includes a layer of inactive cells around
  the active cells.
\item We compute the CR energy density for all involved cells from their CR energy.
\item We use the semi-implicit diffusion solver to update the CR energy density for the cells in our list. Here, every involved interface uses the minimum of the timesteps of the two adjacent cells.
\item We update the CR energy in the involved cells from the new CR energy density.
\end{enumerate}

The resulting semi-implicit local timestepping scheme for diffusion is
fully conservative, but only requires to solve the diffusion problem
on the active cells plus a one-cell boundary layer.

\begin{figure*}
  \centering
  \includegraphics[width=\linewidth]{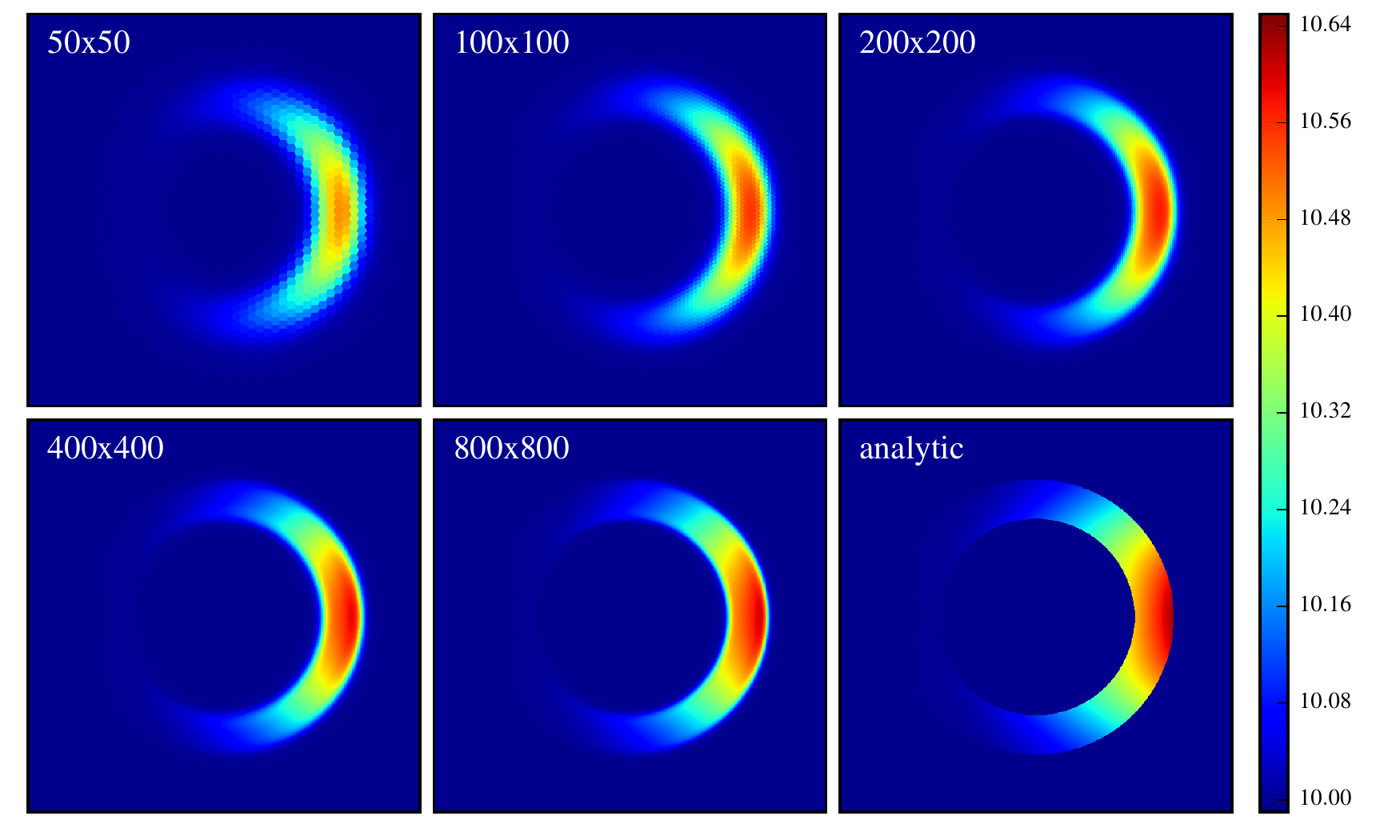}
  \caption{Anisotropic diffusion of a wedge along a circular magnetic
    field in 2D using full normal gradients and global timesteps. The
    panels show the CR energy density at time $t=10$ for different
    resolutions, as labelled, and the analytic solution for
    comparison. The minimum colormap is set to $9.98$ so that any
    values below the background value of $10$ would be clearly visible.}
  \label{fig:AnIsoRingRes}
\end{figure*}

\section{Test problems}
\label{sec:tests}

To understand the accuracy and convergence properties of this scheme
in its different variations we will discuss a set of idealized test
problems. To construct a regular 2D mesh, we take a uniform Cartesian
mesh and move the points in every second row by $0.45\,dx$
horizontally, where $dx$ is the cell size for the given
resolution. Note that the resulting mesh is a regular hexagonal mesh,
but its mesh-generating points are systematically offset from the
centers of mass of the cells by about $1\%$ of the cell radius. We
construct irregular meshes from these regular meshes by adding a
random offset of up to $0.2\,dx$ in both dimensions, mimicking the
typical deviation between mesh-generating points and cell centers in
real problems \citep{Vogelsberger2012}. For all tests except the 2D
blast wave (Sec.~\ref{sec:blast}), we disable hydrodynamics and keep
the mesh fixed. Note that the latter can amplify the overall error on
irregular meshes, as the highly irregular cells stay as they are for
the whole simulation time.

\begin{figure*}
  \centering
  \includegraphics[width=\linewidth]{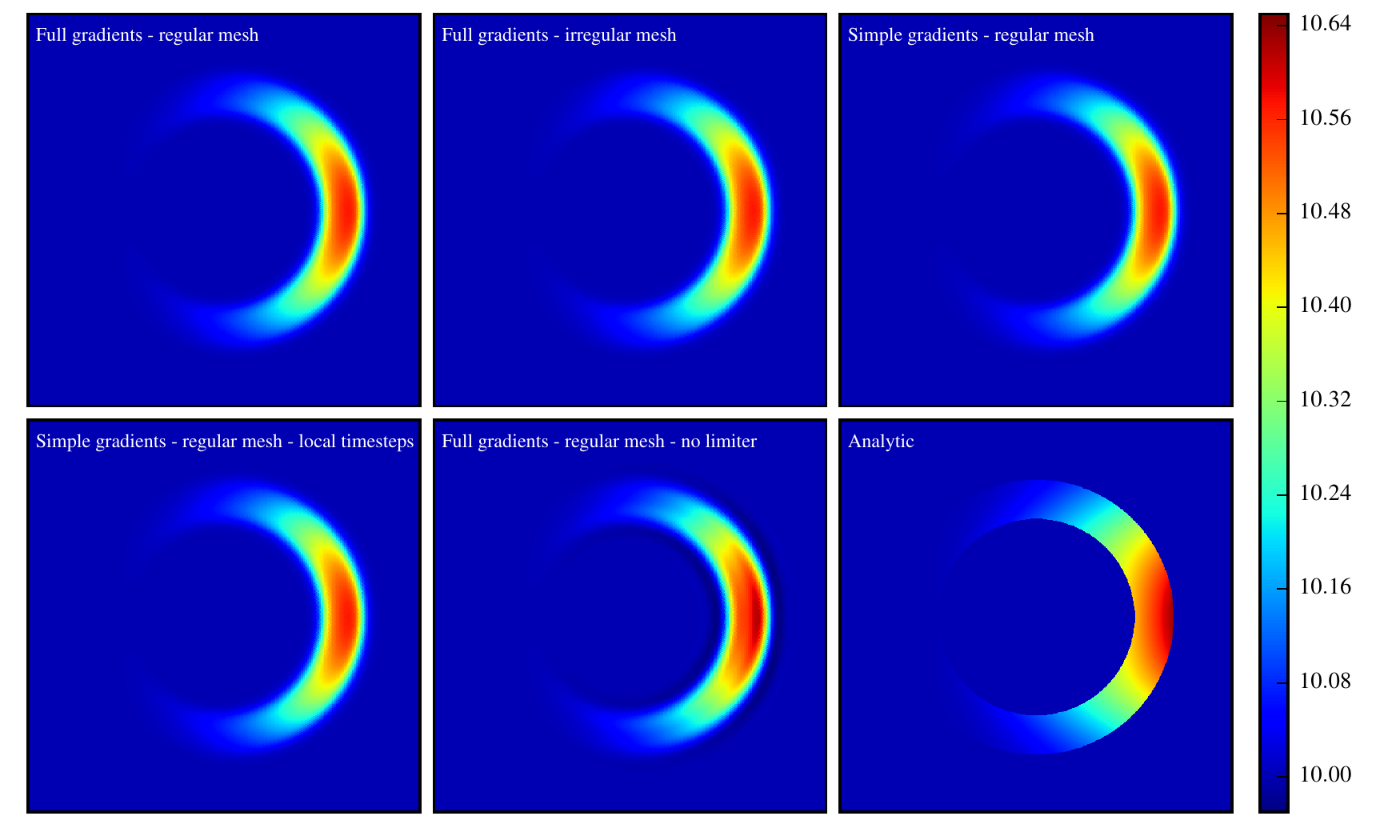}
  \caption{Anisotropic diffusion of a wedge along a circular magnetic
    field in 2D for a resolution of $200\times200$ cells. The panels
    show the CR energy density at time $t=10$ for different methods,
    and the analytic solution for comparison.}
  \label{fig:AnIsoRingMethods}
\end{figure*}

\subsection{Isotropic diffusion of a Gaussian profile}

One of the most straightforward tests for isotropic diffusion is the
evolution of a Gaussian profile over time. On a domain with extension
$[-0.5,0.5]\times [-0.5,0.5]$, we set the CR energy density at the
initial time $t_0=0.1$ to
\begin{equation}
  \label{eq:gauss}
  \eps_\CR \left( \mathbfit{x} \right ) =  1 + \frac{ 10^{-2} }{ 2 \pi D } \times e^{ \frac{-r^2}{2D} },
\end{equation}
where $r = \sqrt{ \mathbfit{x}^2 }$, $D = 2 \kappa_\eps t$, and the
diffusion coefficient is chosen as $\kappa = 0.01$. We initialise all
cells with the value of the CR energy density at their center of
mass. We then advance the simulation from $t_0=0.1$ to $t=0.2$. At
this time, the analytical solution is still given by
Eq.~(\ref{eq:gauss}). The $L^1$ norm measured for different resolutions
on a regular and an irregular mesh is shown in
Fig.~\ref{fig:L1IsoGauss}. There is essentially no dependence of the
accuracy of the solution on the structure of the mesh and the scheme
always converges with second order.

\subsection{Anisotropic diffusion of a wedge in a circular magnetic field}

The anisotropic ring problem has become one of the standard test cases
for anisotropic diffusion. We follow the setup of \citet{Parrish2005}
and \citet{Sharma2007}. It sets the initial CR energy density on
a domain of $[-1,1]^2$ to
\begin{equation}
\eps_\CR  \left( \mathbfit{x} \right ) = 
\begin{cases}
12 & \mbox{if} \ 0.5<r<0.7 \ \mathrm{and} \ \left| \phi \right| < \frac{\displaystyle\pi}{\displaystyle12} \\
10 &  \rm{else}.
\end{cases}
\end{equation}
Here, $r = \sqrt{ \mathbfit{x}^2 }$ is the radial coordinate and $\phi = \mathrm{atan2} \left( y, x \right)$ is the azimuthal coordinate. The magnetic field is given by
\begin{equation}
\begin{aligned}
B_x \left( \mathbfit{x} \right ) &= - \frac{y}{r} \\
B_y \left( \mathbfit{x} \right ) &= + \frac{x}{r}.
\end{aligned}
\end{equation}
Again, we initialise the cells with the values at their centers of mass, and use $\kappa=0.01$.

There are two regimes in which we can easily compute the analytical
solution of the problem. At late times (we usually use $t=200$),
diffusion in the azimuthal directions should have distributed the
energy uniformly in the ring, so that $\eps_\CR = 10 + 1/6$ for
$0.5<r<0.7$ and $\eps_\CR = 10$ everywhere else. The numerical
solution at this time is sensitive to the amount of diffusion
perpendicular to the magnetic field, but is insensitive to errors in
the diffusion speed along magnetic field lines, as any information
along the azimuth has been eliminated as a result of
diffusion. Therefore, it is more demanding to compare the numerical
solution to the analytical solution early on.

At early times (we use $t=10$ here), the energy that starts to diffuse
from the initial wedge along magnetic field rings has not closed the
ring yet. Therefore, we can at every radius treat the problem as a
one-dimensional diffusion problem of a step function. The analytic
solution is given by
\begin{equation}
\eps_\CR  \left( \mathbfit{x} \right ) = 10 + \mathrm{erfc} \left[  \left( \phi + \frac{\pi}{12} \right) \frac{r}{D} \right] - \mathrm{erfc} \left[  \left( \phi - \frac{\pi}{12} \right) \frac{r}{D} \right],
\end{equation}
with $D = \sqrt{ 4 \kappa_\eps t }$ for $0.5<r<0.7$ and $\eps_\CR  \left( \mathbfit{x} \right ) = 10$ everywhere else.

\begin{figure*}
  \centering
  \includegraphics[width=\linewidth]{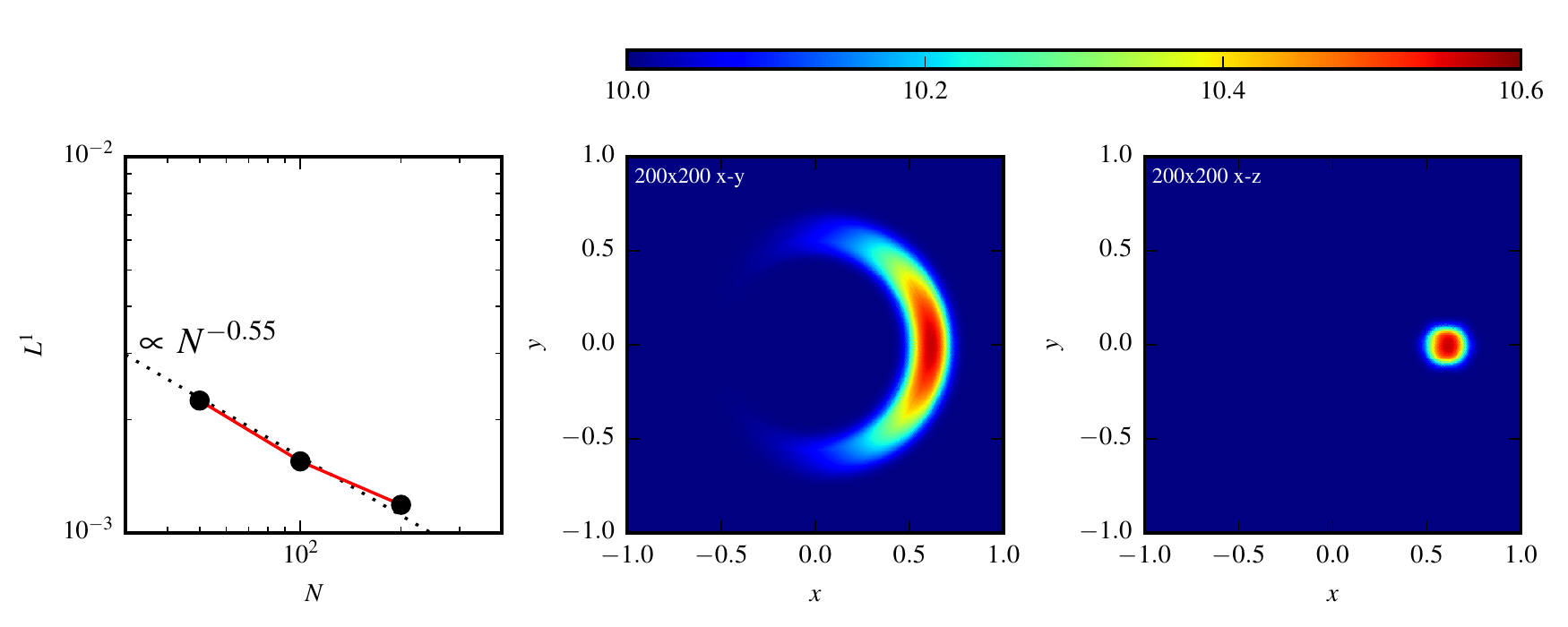}
  \caption{Anisotropic diffusion of a wedge along an azimuthal
    magnetic field in 3D. The left panel shows the $L^1$ norm at time
    $t=10$ for different resolutions. The center and right panels show
    slices of the CR energy density at $t=10$ in the $x-y$ and the
    $x-z$ plane, respectively.}
  \label{fig:AnIsoTorus}
\end{figure*}

The $L^1$ norms for different resolutions and different configurations
of the anisotropic diffusion solver are shown in
Fig.~\ref{fig:L1AnIsoRing}. The convergence rate at early times is
identical for all four runs, independent of the regularity of the
mesh, the use of simple or full normal gradient estimates, and the use
of local or global timesteps. At late times, the only visible
difference is that the error is larger for the run with local
timesteps, but still converges at about the same rate as the other
runs. The convergence rates at $t=10$ ($L^1 \propto N^{-0.55}$) and
$t=200$ ($L^1 \propto N^{-0.7}$) are comparable to or slightly better
than previous results \citep[see,
e.g.][]{Parrish2005,Sharma2007,Kannan2016}. In all runs the minimum CR
energy density is $10$, demonstrating that the limiting is sufficient
to prevent violation of the entropy condition in this problem. Note
also that the convergence rate at $t=200$ is significantly better than
the convergence rate at $t=10$, because the errors of the diffusion
problem along the magnetic field lines are eliminated because of the
memory loss due to diffusion at $t=200$. Moreover, the convergence
rate is significantly worse than first order ($L^1 \propto
N^{-1}$).

Therefore, computing a reference solution at higher resolution and
comparing to this instead of the analytical solution (as done in
\citealt{Hopkins2016DiffAniso}) will significantly overestimate the
convergence rate and is dangerous because it ultimately does not show
that the scheme converges to the correct solution.

\begin{figure}
  \centering
  \includegraphics[width=\linewidth]{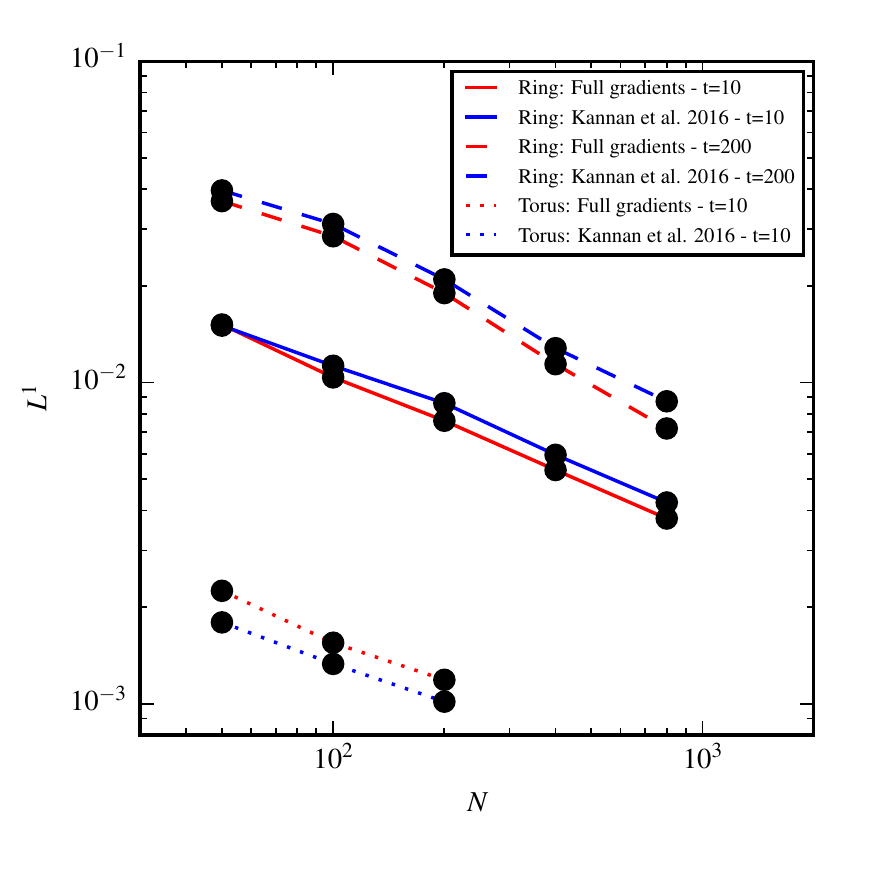}
  \caption{$L^1$ norm for anisotropic diffusion of an initial wedge in
    a circular magnetic field in 2D (ring) and 3D (torus). Solid and
    dashed lines show the $L^1$ norm for the 2D ring problem at $t=10$
    and $t=200$, respectively. The dashed lines show the $L^1$ norm
    for the 3D torus problem at $t=10$. The red lines correspond to
    runs that use the full normal gradient estimate, the blue lines
    represent runs that use the implementation in
    \citet{Kannan2016}. All runs use global timesteps.}
  \label{fig:L1AnIsoRahul}
\end{figure}

Figure~\ref{fig:AnIsoRingRes} shows the CR energy density for the run
with full normal gradients and global timesteps on a regular mesh at
$t=10$ for different resolutions. The main improvements with
resolution are a reduction of diffusion in the radial direction and
therefore a tighter ring, and a less peaked distribution of the CR
energy density at the inner and outer boundaries of the ring. It can
be nicely seen that there are no minima below the background CR energy
density. Figure~\ref{fig:AnIsoRingMethods} shows the result at $t=10$
with a resolution of $200\times200$ cells for different configurations
of the anisotropic diffusion solver. As expected from the very similar
$L^1$ norms, there is essentially no difference in the solutions. For the
mesh setup we use here, there are no problematic corners for the full
normal gradients. In general 3D simulations, the fraction of problematic
corners can increase up to $20\%$. It is interesting to note that deactivating 
the limiter will lead to a measurable reduction of the order or a few percent 
of the CR energy density below the background value at the borders of the ring.

\begin{figure*}
  \centering
  \includegraphics[width=\linewidth]{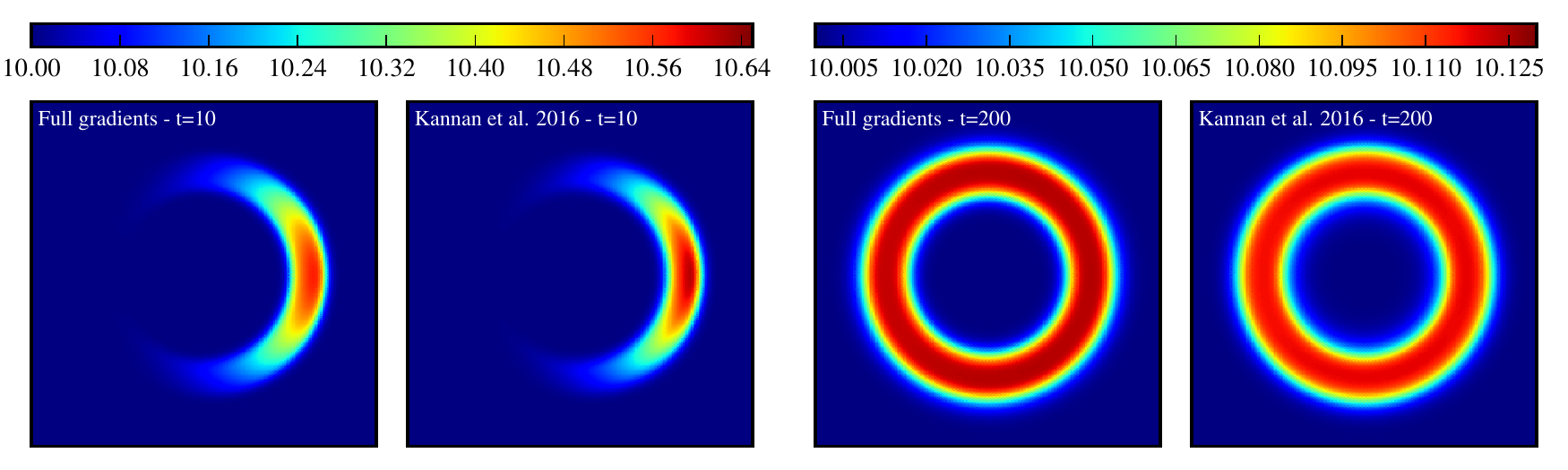}
  \caption{Anisotropic diffusion of an initial wedge in a circular
    magnetic field in 2D for a resolution of $200\times200$ cells. The
    pair of panels on the left shows the CR energy density at $t=10$,
    the pair of panels on the right at $t=200$. The left panel of each
    pair shows a simulation that uses the full normal gradient
    estimate, the right panel of each pair shows a run that uses the
    implementation in \citet{Kannan2016}. All simulations employ global
    timesteps.}
  \label{fig:RingSlicesRahul}
\end{figure*}

Although we have shown that our scheme works well in 2D, a 3D test is also necessary, because the estimate of the gradient is significantly more complex in 3D. To this end, we generalise the anisotropic ring problem to 3D. Our initial wedge is now defined on the domain $[-0.5,0.5]^3$ by
\begin{equation}
\eps_\CR  \left( \mathbfit{x} \right ) = 
\begin{cases}
  12 & \!\!\mbox{if} \ \left[(0.6 - r)^2 + z^2 - 10^{-2} < 0\right]  \land  \left(
  \left| \phi \right| < \frac{\pi}{12}\right), \\
  10 & \!\!\rm{else}.
\end{cases}
\end{equation}
The $z$-component of the magnetic field is set to zero. The
simulations are run on an irregular 3D mesh, starting from a hexagonal
close-packed mesh and then randomly displacing mesh-generating points
in the same way as done in 2D. The analytical solution at $t=10$ can
again be computed by splitting the problem into 1D problems in
azimuth.

The $L^1$ norm at $t=10$ and slices of the $x-y$ plane and the $x-z$
plane through the center are shown in Fig.~\ref{fig:AnIsoTorus}. The
convergence rate is essentially identical to the 2D problem. While the
slice through the $x-y$ plane looks very similar to the 2D
simulations, the slice through the $x-z$ plane shows that numerical
diffusion in the radial and vertical directions is approximately of
equal size.

\subsection{Comparing the anisotropic diffusion solver with the method
  of \citet{Kannan2016} }

While developing the scheme presented in this paper, our collaboration
worked in parallel on a fundamentally different approach to
anisotropic diffusion on unstructured meshes, which has recently been
discussed in detail in \citet{Kannan2016}. Here we briefly compare the
two algorithms on the 2D ring and the 3D torus problems on a slightly
disturbed mesh. The two main differences between the methods are that
the scheme presented in \citet{Kannan2016} operates on the positions
of the mesh-generating points in contrast to the scheme presented here
that uses the centers of mass of the cells, and the size of the
stencil needed differs when the time integration is generalised to
individual timesteps. 

In the scheme presented here, we need to include only the direct
neighbours of active cells to compute the fluxes over all active
interfaces which allows for the straightforward extension to local
timesteps described above. In contrast, for the scheme presented in
\citet{Kannan2016}, we need to include a second layer of cells around
active cells. However, this obviously does not make a difference for
time integrations schemes that only work on global timesteps.

The $L^1$ norm at $t=10$ and $t=200$ for the 2D ring problem and the
$L^1$ norm at $t=10$ for the 3D torus problem for both schemes are
shown in Fig.~\ref{fig:L1AnIsoRahul}. Both schemes converge
essentially at the same rate, only the normalisation is slightly
smaller in 2D and slightly larger in 3D for the method presented in
this paper. The results for the 2D ring problem for the simulations
with both schemes for a resolution of $200\times200$ cells at $t=10$
and $t=200$ are shown in Fig.~\ref{fig:RingSlicesRahul}. While they
are overall very similar, the errors in the two schemes are slightly
different. At $t=10$, the maximum value is slightly lower in our
scheme compared to the scheme presented in \citet{Kannan2016}. At the
same time, the solution for the method presented in \citet{Kannan2016}
is more radially peaked towards the center of the ring. At $t=200$,
the solutions are again very similar with a slightly higher peak value
for the scheme presented here, which translates into a slightly
smaller amount of numerical diffusion in the radial direction. We
conclude that both schemes perform almost identically well and the
differences are only in the details.

\begin{figure}
  \centering
  \includegraphics[width=\linewidth]{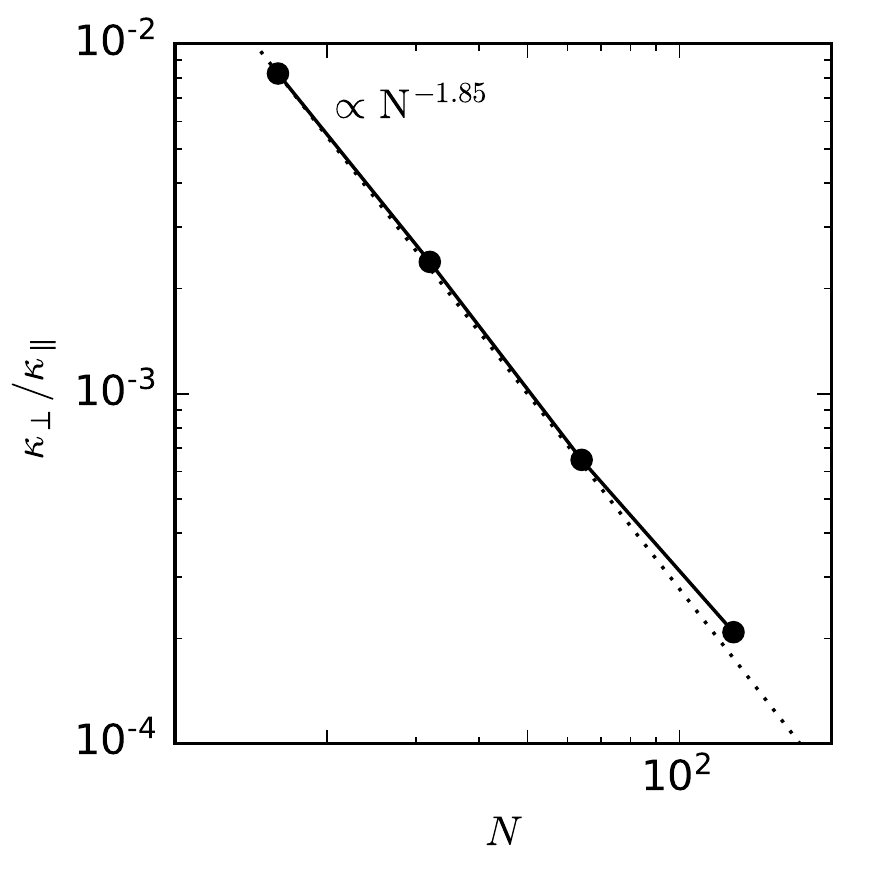}
  \caption{Ratio of the perpendicular numerical diffusion coefficient
    to the parallel diffusion coefficient for the Sovinec test problem
    as a function of resolution.}
  \label{fig:sovinec}
\end{figure}

\subsection{Measuring the perpendicular numerical diffusion coefficient with the Sovinec test}

It is possible to directly measure the effective perpendicular
diffusion coefficient that is entirely numerical in our simulations
\citep{Sovinec2004}. We again use the same regular mesh we used for
the 2D ring problem on a domain of $[-0.5,0.5]^2$, deactivate
hydrodynamics, and keep the mesh fixed. We initialise the CR energy
density with zero and the purely azimuthal magnetic field is given by
\begin{equation}
\begin{aligned}
B_{x} \left( \mathbfit{x} \right ) &= +\cos{ (x \pi) } \sin{ (y \pi) }, \\
B_{y} \left( \mathbfit{x} \right ) &= - \sin{ (x \pi) } \cos{ (y \pi) }.
\end{aligned}
\end{equation}
CR energy is injected at a constant rate following the source term
\begin{equation}
\left. \frac{\partial \eps_\CR \left( x, y \right) }{\partial t} \right|_\mathrm{src} = 2 \pi^2 \cos{ (x \pi) } \cos{ (y \pi) }.
\end{equation}

\begin{figure*}
  \centering
  \includegraphics[width=\linewidth]{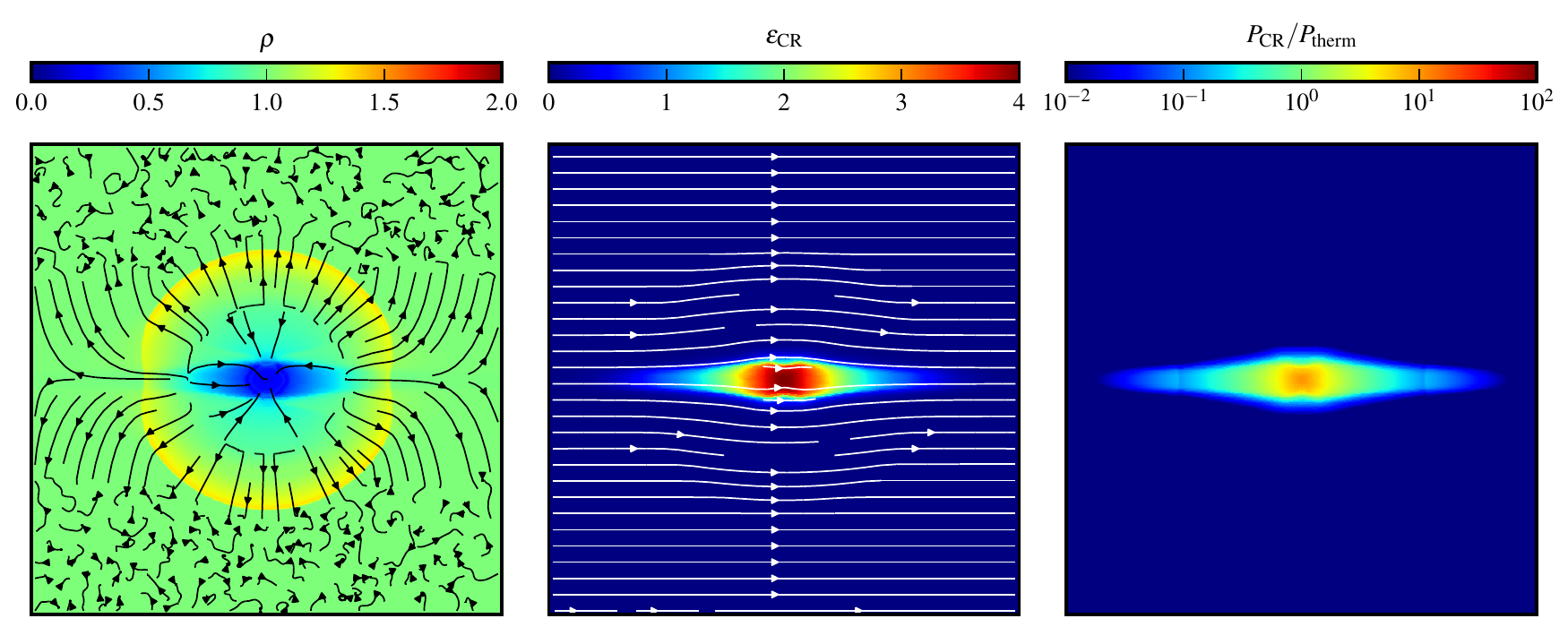}
  \caption{Slices of a 2D CR-driven blast wave with anisotropic CR
    diffusion. The left panel shows the gas density with velocity
    streamlines. The middle panel shows the CR energy density with
    magnetic field lines. The right panel shows the ratio between CR
    and thermal pressure. The simulation has been run with
    $512\times512$ cells.}
  \label{fig:blast2D}
\end{figure*}

We employ a parallel diffusion coefficient of $\kappa_\parallel = 1$
and a perpendicular diffusion coefficient of $\kappa_\perp = 0$. We
use outflow boundary conditions and fix the value of the CR energy
density outside the computational domain to zero. Without numerical
diffusion in the radial direction the CR energy in the computational
domain should increase to infinity. However, because there is
non-negligible numerical diffusion in the radial direction the CR energy
density in the computational domain will eventually reach an
equilibrium when the radial gradient is steep enough such that the
radial diffusion out of the box balances the injection by the source
term. Once equilibrium is reached, the effective numerical diffusion
coefficient in the radial direction is directly given by the inverse
of the CR energy density in the center of the computational domain
\citep{Sovinec2004},
\begin{equation}
\kappa_{\perp,\mathrm{num}} = \eps_\CR \left( 0, 0 \right)^{-1}.
\end{equation}

\begin{figure}
  \centering
  \includegraphics[width=\linewidth]{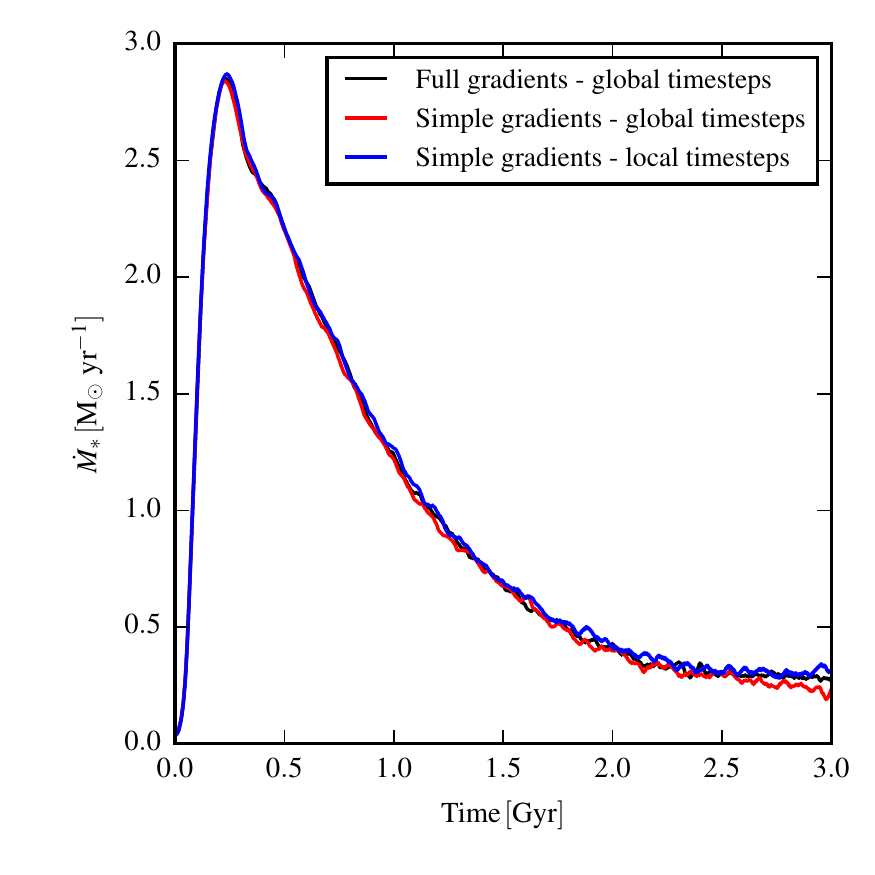}
  \caption{Time evolution of the star formation rate for the isolated
    disk for different configurations of the anisotropic diffusion
    solver.}
  \label{fig:DiskTimeEvolSfr}
\end{figure}

\begin{figure}
  \centering
  \includegraphics[width=\linewidth]{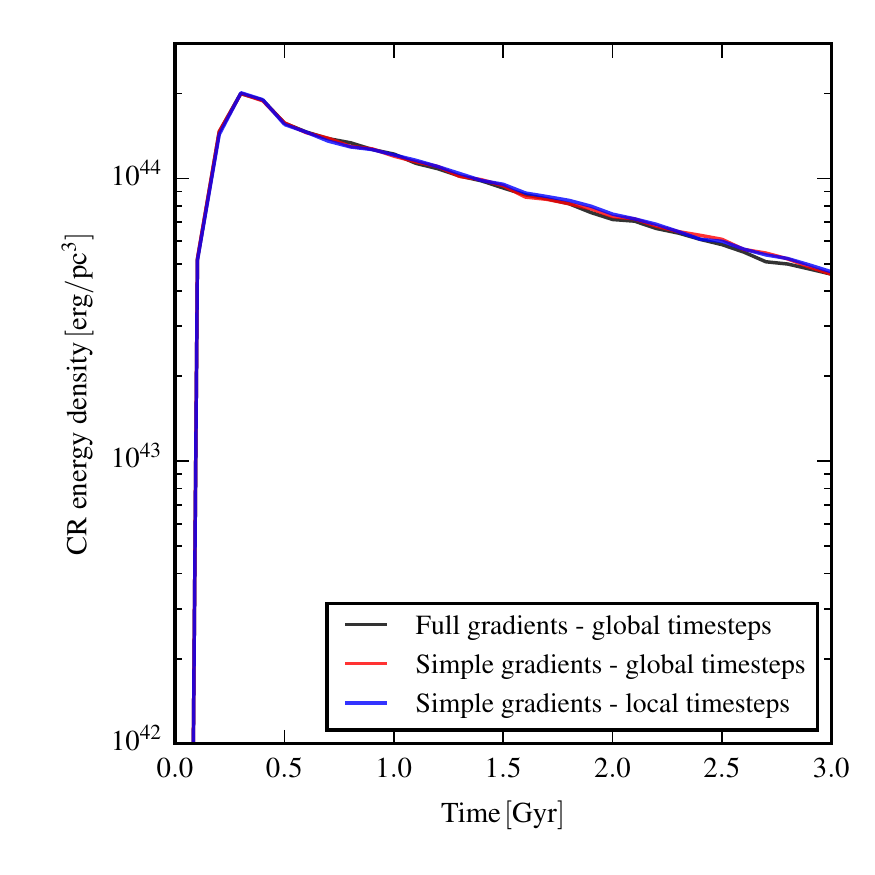}
  \caption{Time evolution of average CR energy density in a cylinder
    with a radius of $10$~kpc and a height of $1$~kpc, centered on the
    disk for different configurations of the anisotropic diffusion
    solver.}
  \label{fig:DiskTimeEvolEcr}
\end{figure}

\begin{figure*}
  \centering
  \includegraphics[width=\linewidth]{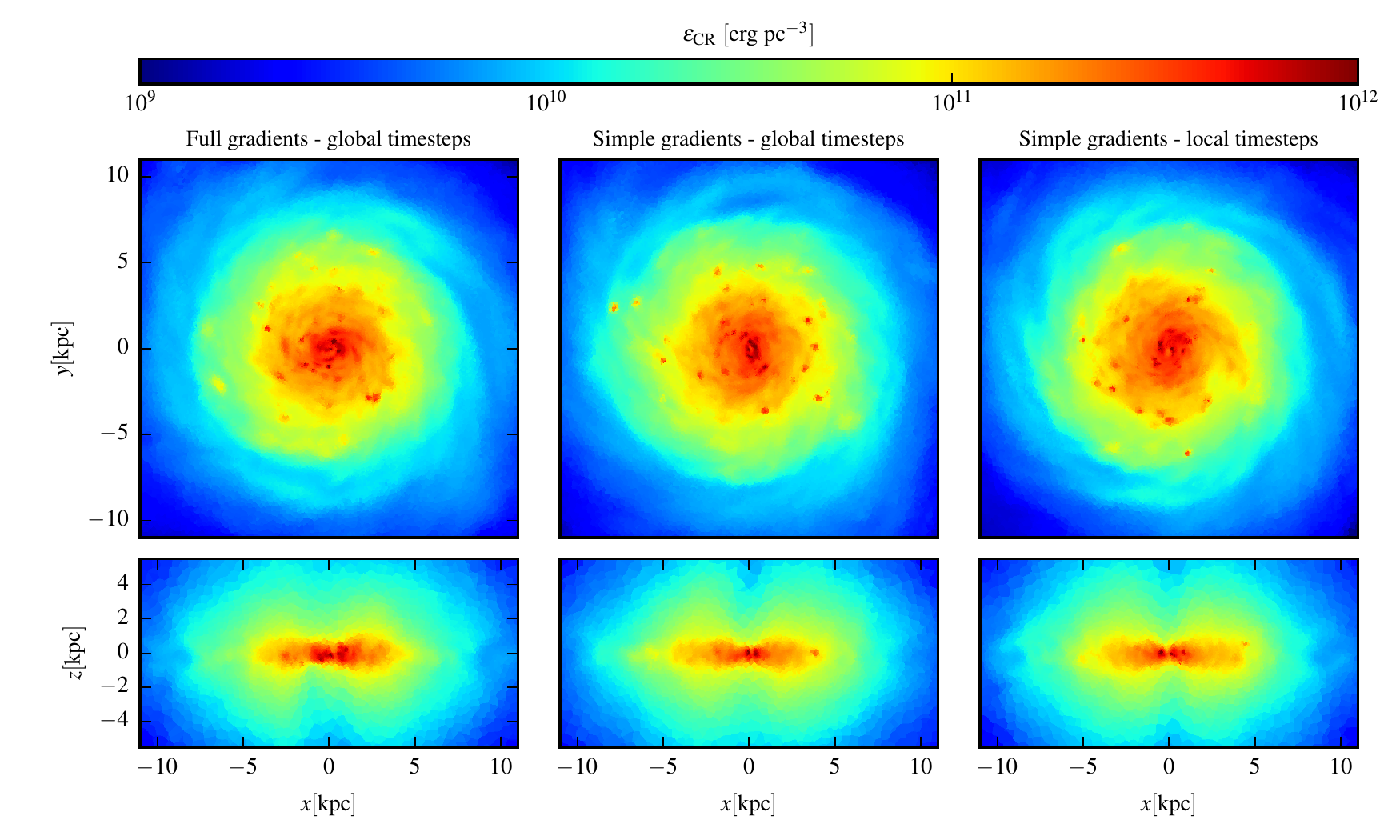}
  \caption{Slices of the CR energy density at time $t=1.5$~Gyrs. The
    columns show simulations with different configurations of the
    anisotropic diffusion solver. Top and bottom rows show face-on and
    edge-on slices through the center of the disk, respectively.}
  \label{fig:DiskSlices}
\end{figure*}

\begin{figure*}
  \centering
  \includegraphics[width=\linewidth]{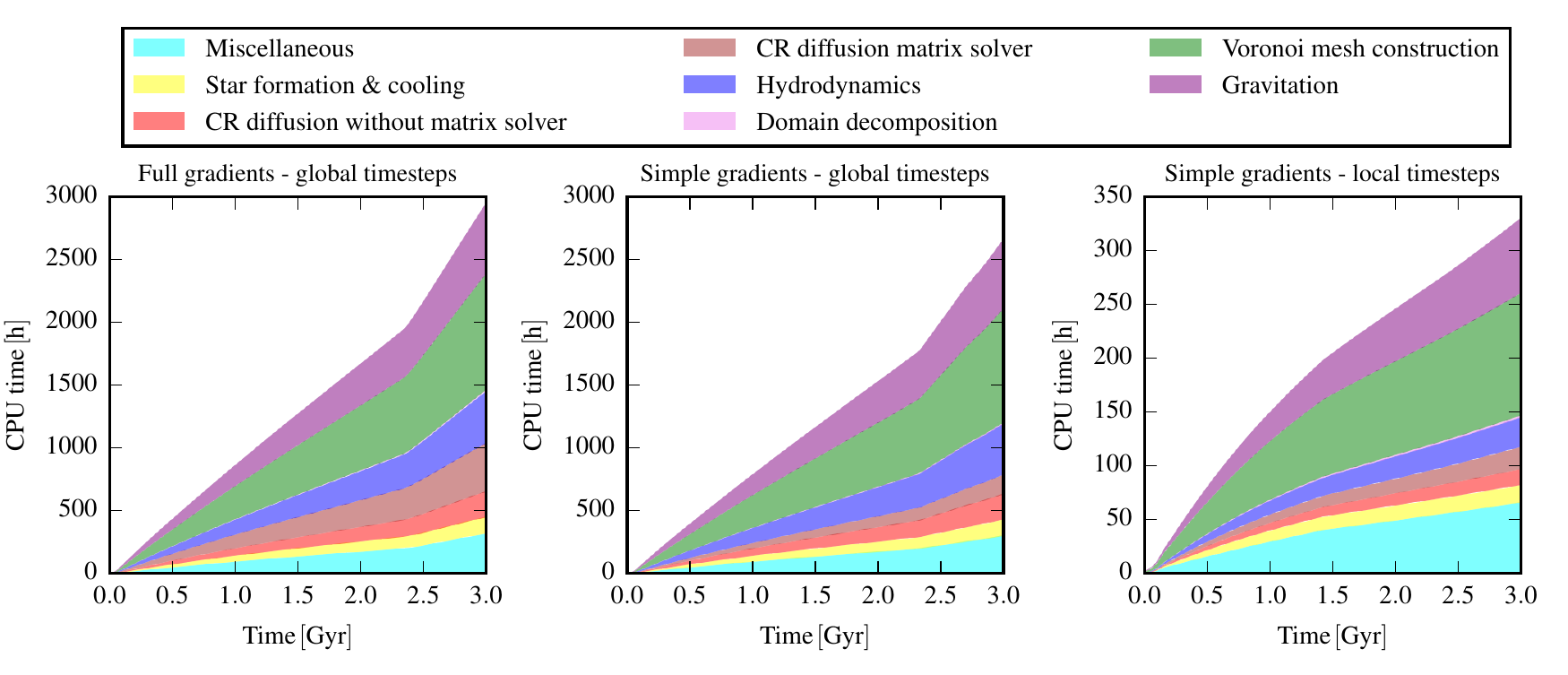}
  \caption{CPU time needed to run the isolated disk simulations to
    $t=3$~Gyrs for different configurations of the anisotropic
    diffusion solver. Note the different scale on the CPU time axis in
    the panel on the right-hand side.}
  \label{fig:DiskCPU}
\end{figure*}

The ratio $\kappa_{\perp,\mathrm{num}} / \kappa_\parallel$ for
different resolutions is shown in Fig.~\ref{fig:sovinec}. The
effective perpendicular diffusion coefficient decreases almost with
second order. Moreover, even for a resolution of only $16^2$ cells it
is already smaller than $10^{-2}$ and decreases rapidly. This result
is comparable or even slightly better than previous implementations of
anisotropic diffusion
\citep{Parrish2005,Sharma2007,Kannan2016}. Therefore, we conclude that
our implementation is able to guarantee a negligible amount of
perpendicular numerical diffusion in practical problems.

\subsection{2D blast wave with anisotropic diffusion}
\label{sec:blast}

To test the anisotropic diffusion solver fully coupled with ideal MHD,
we simulate a CR driven point explosion in 2D. We start with a regular
mesh on a domain $[-0.5,0.5]^2$ with a homogeneous density of
$\rho = 1$, a homogeneous specific thermal energy of $u = 2.5$, and a
uniform magnetic field along the $x$-axis of $B_{x} = 1$. We model the
CRs with a constant $\gamma_\CR=4/3$ \citep{Pfrommer2016} and set the
initial CR energy to $\eps_\CR = 100$ for $r < 0.02$ and to zero at
larger radii. We set $\gamma_\mathrm{gas}=5/3$ and the kinetic
energy-weighted anisotropic CR diffusion coefficient to
$\kappa_\eps=0.1$. We employ the new second order hydro scheme in
\textsc{arepo} \citep{Pakmor2016} and use its standard ideal MHD
implementation \citep{Pakmor2011,Pakmor2013}.

Figure~\ref{fig:blast2D} shows the simulation at $t=0.1$. The central
injection of CR energy initially drives a spherical shock wave that is
modified as CRs are able to escape along the magnetic field lines
along the $x$-axis. There is no diffusion of CRs in the $y$-direction,
therefore the shock wave in this direction behaves essentially like in
a Sedov solution without diffusion. In particular, the gas in front of
the shock is unperturbed. Close to the center the magnetic field lines
are somewhat advected in the $y$-direction because of adiabatic
expansion, but the $x$-component of the magnetic field still
dominates. In contrast, CRs are able to diffuse freely in the
$x$-direction, which leads to an enhancement of CRs ahead of the shock
that already preconditions the velocity field ahead of the shock.

As in the pure diffusion test problems, there is essentially no
diffusion of CRs perpendicular to the magnetic field lines, and the
broadening of the CR energy density in the $y$-direction in the center
is purely a result of advection due to adiabatic expansion.

\begin{figure*}
  \centering
  \includegraphics[width=\linewidth]{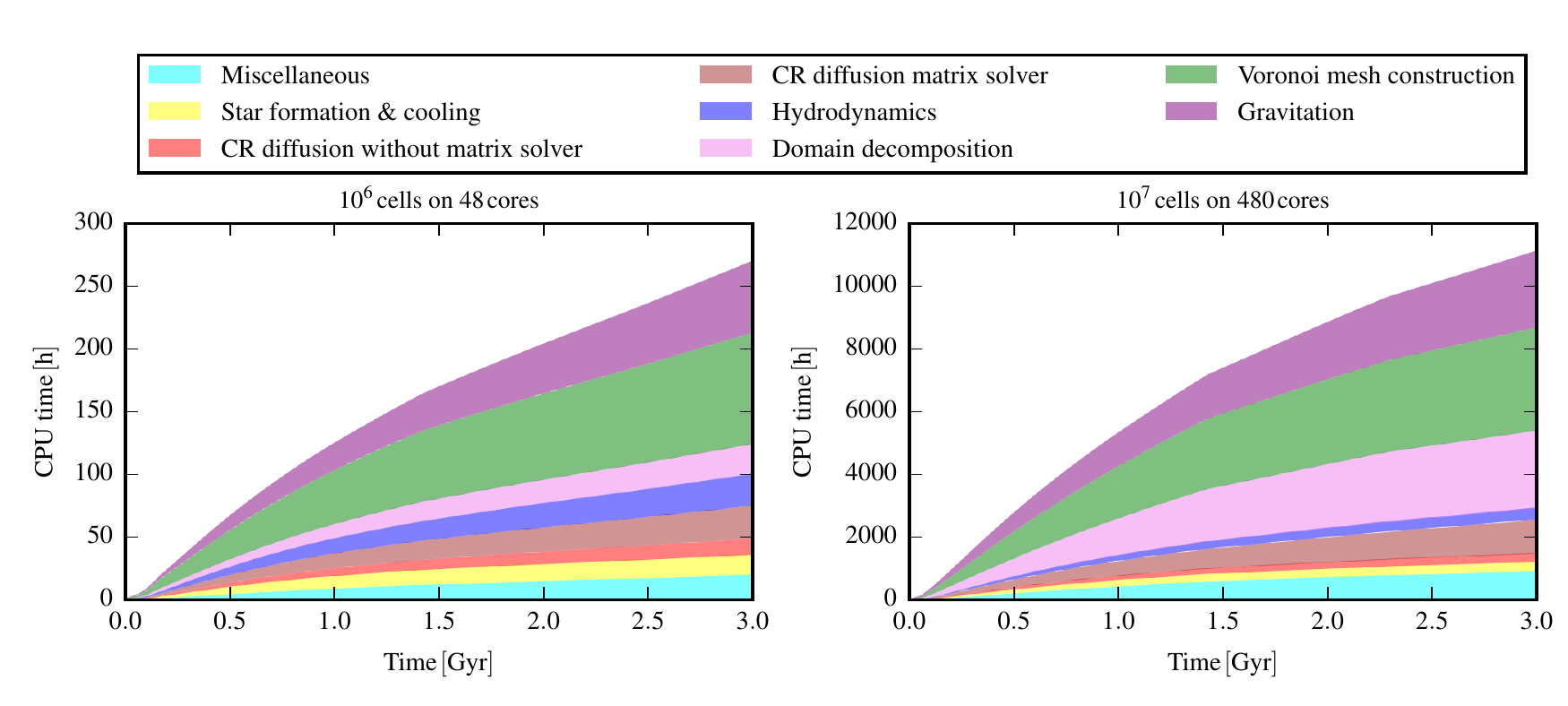}
  \caption{CPU time needed to run the isolated disk simulations to
    $t=3$~Gyrs for different resolutions and number of CPUs.}
  \label{fig:DiskScaling}
\end{figure*}

\section{Anisotropic CR diffusion in isolated disk galaxies}
\label{sec:galaxy}

One of the most interesting applications of anisotropic CR diffusion
is the field of galaxy evolution, in particular as a driving mechanism
for galactic winds \citep{Uhlig2012, Booth2013, Salem2014, Salem2014b,
  Ruszkowski2016}. Here we concentrate on testing our methodology by
looking for differences caused by different configurations of the
anisotropic diffusion solver in simulations of the formation and
evolution of an isolated disk galaxy. For a detailed analysis of the
physical implications of these simulations we refer to
\citet{Pakmor2016b}.

Our setup is close to one previously used in \citet{Pakmor2013}. A
static background dark matter halo that follows a spherical
NFW-profile \citep{NFW1} with a concentration of $7.2$ and a total
mass of $10^{11}\,\mathrm{M_\odot}$ is filled with gas in hydrostatic
equilibrium (yielding a baryon fraction of $0.17$ in the halo) and
made to rotate around the $z$-axis with a spin parameter of
$\lambda=0.05$. We also add a uniform magnetic seed field along the
$x$-axis with a strength of $10^{-10}\mathrm{G}$. We discretise the
gas into $10^6$ cells of equal mass with a mass of
$2\times10^4\,\mathrm{M_\odot}$ and use the standard configuration for
refinement and mesh regularization
\citep{Vogelsberger2012,Pakmor2016}. The interstellar medium is
modelled by an effective equation of state \citep{Springel2003}. Star
formation is handled probabilistically with a star formation rate that
only depends on the local gas density \citep{Springel2003}.

We model the CRs in a two-fluid approximation with a constant
adiabatic index of $4/3$ \citep{Pfrommer2016}. We initially set the CR
energy density in all cells to zero, and whenever a star particle is
created we inject $10^{48}$~erg of CR energy per solar mass of formed
stars immediately into the neighbouring cells. We employ a kinetic
energy-weighted diffusion coefficient for anisotropic CR diffusion of
$\kappa_\eps=10^{28}\mathrm{cm^2\,s^{-1}}$ and model CR cooling as
described in \citet{Pfrommer2016}.

In Figs.~\ref{fig:DiskTimeEvolSfr} and \ref{fig:DiskTimeEvolEcr}, we
show the time evolution of the total star formation rate and the
average CR energy density in a cylinder of radius $10$~kpc and height
$1$~kpc centered on the disk, for different configurations of the
anisotropic diffusion solver in otherwise identical runs. Both
quantities are sensitive probes of the amount of CR diffusion out of
the disk. Neither the change from the full normal gradient estimate to
the simple normal gradient estimate, nor using local timesteps instead
of global timesteps for the integration of the anisotropic diffusion
problem have any impact on the results beyond the intrinsic noise
induced by the probabilistic description of star formation.

This result is reinforced by Fig.~\ref{fig:DiskSlices}, which shows
slices of the CR energy density through the disk at $t=15$~Gyrs for
three different configurations of the anisotropic diffusion
solver. While there are differences in detail as the CR energy is
injected in different places in the disk in the different runs, the
overall CR energy distribution is very similar. The use of a constant
diffusion coefficient quickly smoothes out any discontinuities in
$\eps_\CR$ as soon as anisotropic diffusion along the azimuthal
magnetic field is efficient.

However, the different configurations of the anisotropic diffusion
solver lead to very different runtimes for the same problem as shown
in Fig.~\ref{fig:DiskCPU}, even though the result is essentially
identical as discussed above. In the most conservative configuration
that employs global timesteps and the full normal gradient estimate,
the CR diffusion solver takes about $20\%$ of the total
runtime. Replacing the full normal gradient estimate with the simple
normal gradient estimate reduces the fraction of the runtime used by
the CR diffusion to about $13\%$. While these numbers clearly depend
on the problem and more importantly on the resolution (the fraction of
the total runtime used by the CR diffusion solver generally increases
with resolution, as it scales worse than the gravity and hydrodynamics
solvers), there is a significant speedup associated with using the
simple normal gradient estimate.

A much larger speedup is achieved when changing from global to
individual local timesteps, because the number of cells that requires a
small timestep is often negligible compared to the total number of
cells. As shown in Fig.~\ref{fig:DiskCPU}, the total runtime drops by
about a factor of $10$ while the fraction of the total runtime spent
in the CR diffusion solver is only $10\%$. This clearly demonstrates
again the importance of being able to use individual local timesteps to
efficiently run simulations of galaxy formation. In fact, this is
extremely critical for being able to simulate the formation and
evolution of galaxies in the full cosmological framework at high
resolution.

For all configurations, about half of the computational time of
the diffusion problem is spent in the matrix solver. We
expect this part of the calculation to be comparable in cost to similar solvers on
Cartesian meshes \citep[see, e.g.][]{Parrish2005,Sharma2007}.
The remaining parts of the diffusion solver are
dominated by the computation of the coefficients for the gradient estimates
and these calculations will be more expensive on our unstructured mesh. However,
owing to the large speed-up obtained from using local timesteps, the
diffusion solver only uses a small fraction of the overall runtime.

The weak scaling of the isolated galaxy simulation done with \textsc{Arepo}, 
including the semi-implicit diffusion solver, is explored in Fig.~\ref{fig:DiskScaling} 
with local timesteps and the full normal
gradient estimate configuration for the diffusion solver. The run
with $10^7$ cells requires about $40$ times more CPU time compared
to the run with $10^6$ cells. The factor of $40$ is a result of a
factor of $10$ for the increased number of resolution elements; a
factor of $\sqrt[3]{10} \approx 2.2$ from the reduction in the timestep from
smaller cell sizes; and a factor $\leq 2$ owing to the larger
peak densities reached in the center of the galaxy. Interestingly,
the CPU fraction spent in the diffusion solver is very similar for
both runs, i.e.~the weak scaling of the diffusion solver is
comparable to the other parts of the code that dominate the runtime
(gravity and mesh construction). The domain decomposition is the
only part of the code that significantly increases its fraction of the
total CPU time. However, this inefficiency could be alleviated by reducing the
frequency of domain decompositions, which has not been attempted
here. The good scaling of the diffusion solver suggests that it can be
readily applied to large cosmological simulations in its current
form.

\section{Summary}
\label{sec:summary}

We presented the formulation and implementation of a new numerical
approach for treating the isotropic and anisotropic diffusion problem
on unstructured, Voronoi meshes as used in the hydrodynamical {\small
  AREPO} code. The main features of our solver are:
\begin{itemize}
\item It is fully conservative.
\item It does not violate the entropy condition $\Delta S>0$ in any significant way.
\item It provides a semi-implicit time integration scheme with individual timesteps that only requires the solution of a single linear system of equations per timestep.
\item Its convergence rate for isotropic diffusion is fully second order.
\item For anisotropic diffusion, it has a comparable or slightly
  better convergence rate compared to previous implementations on
  structured and unstructured meshes.
\end{itemize}

The combination of these properties allows us to efficiently use the
anisotropic diffusion solver in simulations of structure formation as
well as galaxy formation and evolution without the need to make any
compromises in terms of accuracy, conservation of energy, or
violations of the entropy condition. The new method is thus enabling
realistic treatments of cosmic ray physics in (computationally
expensive) studies of galaxy formation, which is a particularly timely
problem due to the relevance of cosmic rays for feedback processes. We
demonstrate this in first science applications of our new methodology
\citep{Pakmor2016b, Simpson2016}, and expect that our new approach
will be very useful in future work that aims for a comprehensive
modelling of galaxy formation physics.

\section*{Acknowledgements}
    
This work has been supported by the European Research Council under
ERC-StG grant EXAGAL-308037, ERC-CoG grant CRAGSMAN-646955, and by the
Klaus Tschira Foundation. VS acknowledges support through subproject
EXAMAG of the Priority Programme 1648 `Software for Exascale
Computing' of the German Science Foundation.
    
\bibliographystyle{mnras}

\begin{thebibliography}{}
\makeatletter
\relax
\def\mn@urlcharsother{\let\do\@makeother \do\$\do\&\do\#\do\^\do\_\do\%\do\~}
\def\mn@doi{\begingroup\mn@urlcharsother \@ifnextchar [ {\mn@doi@}
  {\mn@doi@[]}}
\def\mn@doi@[#1]#2{\def\@tempa{#1}\ifx\@tempa\@empty \href
  {http://dx.doi.org/#2} {doi:#2}\else \href {http://dx.doi.org/#2} {#1}\fi
  \endgroup}
\def\mn@eprint#1#2{\mn@eprint@#1:#2::\@nil}
\def\mn@eprint@arXiv#1{\href {http://arxiv.org/abs/#1} {{\tt arXiv:#1}}}
\def\mn@eprint@dblp#1{\href {http://dblp.uni-trier.de/rec/bibtex/#1.xml}
  {dblp:#1}}
\def\mn@eprint@#1:#2:#3:#4\@nil{\def\@tempa {#1}\def\@tempb {#2}\def\@tempc
  {#3}\ifx \@tempc \@empty \let \@tempc \@tempb \let \@tempb \@tempa \fi \ifx
  \@tempb \@empty \def\@tempb {arXiv}\fi \@ifundefined
  {mn@eprint@\@tempb}{\@tempb:\@tempc}{\expandafter \expandafter \csname
  mn@eprint@\@tempb\endcsname \expandafter{\@tempc}}}

\bibitem[\protect\citeauthoryear{{Blandford} \& {Eichler}}{{Blandford} \&
  {Eichler}}{1987}]{Blandford1987}
{Blandford} R.,  {Eichler} D.,  1987, \physrep, \href
  {http://esoads.eso.org/cgi-bin/nph-bib_query?bibcode=1987PhR...154....1B&db_key=AST}
  {154, 1}

\bibitem[\protect\citeauthoryear{{Booth}, {Agertz}, {Kravtsov}  \&
  {Gnedin}}{{Booth} et~al.}{2013}]{Booth2013}
{Booth} C.~M.,  {Agertz} O.,  {Kravtsov} A.~V.,   {Gnedin} N.~Y.,  2013,
  \mn@doi [\apjl] {10.1088/2041-8205/777/1/L16}, \href
  {http://adsabs.harvard.edu/abs/2013ApJ...777L..16B} {777, L16}

\bibitem[\protect\citeauthoryear{{Boulares} \& {Cox}}{{Boulares} \&
  {Cox}}{1990}]{Boulares1990}
{Boulares} A.,  {Cox} D.~P.,  1990, \mn@doi [\apj] {10.1086/169509}, \href
  {http://adsabs.harvard.edu/abs/1990ApJ...365..544B} {365, 544}

\bibitem[\protect\citeauthoryear{{Brunetti} \& {Lazarian}}{{Brunetti} \&
  {Lazarian}}{2007}]{Brunetti2007}
{Brunetti} G.,  {Lazarian} A.,  2007, \mn@doi [\mnras]
  {10.1111/j.1365-2966.2007.11771.x}, \href
  {http://adsabs.harvard.edu/abs/2007MNRAS.378..245B} {378, 245}

\bibitem[\protect\citeauthoryear{{Chandran}}{{Chandran}}{2000}]{Chandran2000}
{Chandran} B.~D.~G.,  2000, \mn@doi [Physical Review Letters]
  {10.1103/PhysRevLett.85.4656}, \href
  {http://esoads.eso.org/abs/2000PhRvL..85.4656C} {85, 4656}

\bibitem[\protect\citeauthoryear{{Desiati} \& {Zweibel}}{{Desiati} \&
  {Zweibel}}{2014}]{Desiati2014}
{Desiati} P.,  {Zweibel} E.~G.,  2014, \mn@doi [\apj]
  {10.1088/0004-637X/791/1/51}, \href
  {http://adsabs.harvard.edu/abs/2014ApJ...791...51D} {791, 51}

\bibitem[\protect\citeauthoryear{{Dubois} \& {Commer{\c c}on}}{{Dubois} \&
  {Commer{\c c}on}}{2016}]{Dubois2016}
{Dubois} Y.,  {Commer{\c c}on} B.,  2016, \mn@doi [\aap]
  {10.1051/0004-6361/201527126}, \href
  {http://adsabs.harvard.edu/abs/2016A%26A...585A.138D} {585, A138}

\bibitem[\protect\citeauthoryear{Falgout \& Yang}{Falgout \&
  Yang}{2002}]{HYPRE}
Falgout R.~D.,  Yang U.~M.,  2002, in Proceedings of the International
  Conference on Computational Science-Part III. ICCS '02.
Springer-Verlag, London, UK, UK, pp 632--641, \url
  {http://dl.acm.org/citation.cfm?id=645459.653635}

\bibitem[\protect\citeauthoryear{{Girichidis} et~al.,}{{Girichidis}
  et~al.}{2016}]{Girichidis2016}
{Girichidis} P.,  et~al., 2016, \mn@doi [\apjl] {10.3847/2041-8205/816/2/L19},
  \href {http://adsabs.harvard.edu/abs/2016ApJ...816L..19G} {816, L19}

\bibitem[\protect\citeauthoryear{{G{\"u}nter}, {Yu}, {Kr{\"u}ger}  \&
  {Lackner}}{{G{\"u}nter} et~al.}{2005}]{Guenter2005}
{G{\"u}nter} S.,  {Yu} Q.,  {Kr{\"u}ger} J.,   {Lackner} K.,  2005, \mn@doi
  [Journal of Computational Physics] {10.1016/j.jcp.2005.03.021}, \href
  {http://adsabs.harvard.edu/abs/2005JCoPh.209..354G} {209, 354}

\bibitem[\protect\citeauthoryear{{Hanasz}, {Lesch}, {Naab}, {Gawryszczak},
  {Kowalik}  \& {W{\'o}lta{\'n}ski}}{{Hanasz} et~al.}{2013}]{Hanasz2013}
{Hanasz} M.,  {Lesch} H.,  {Naab} T.,  {Gawryszczak} A.,  {Kowalik} K.,
  {W{\'o}lta{\'n}ski} D.,  2013, \mn@doi [\apjl] {10.1088/2041-8205/777/2/L38},
  \href {http://adsabs.harvard.edu/abs/2013ApJ...777L..38H} {777, L38}

\bibitem[\protect\citeauthoryear{Henson \& Yang}{Henson \&
  Yang}{2002}]{Henson2002}
Henson V.~E.,  Yang U.~M.,  2002, \mn@doi [Applied Numerical Mathematics]
  {http://dx.doi.org/10.1016/S0168-9274(01)00115-5}, 41, 155

\bibitem[\protect\citeauthoryear{{Hopkins}}{{Hopkins}}{2016}]{Hopkins2016DiffAniso}
{Hopkins} P.~F.,  2016, preprint, \href
  {http://adsabs.harvard.edu/abs/2016arXiv160207703H} {} (\mn@eprint {arXiv}
  {1602.07703})

\bibitem[\protect\citeauthoryear{{Kannan}, {Springel}, {Pakmor}, {Marinacci}
  \& {Vogelsberger}}{{Kannan} et~al.}{2016}]{Kannan2016}
{Kannan} R.,  {Springel} V.,  {Pakmor} R.,  {Marinacci} F.,   {Vogelsberger}
  M.,  2016, \mn@doi [\mnras] {10.1093/mnras/stw294}, \href
  {http://adsabs.harvard.edu/abs/2016MNRAS.458..410K} {458, 410}

\bibitem[\protect\citeauthoryear{{Kulsrud}}{{Kulsrud}}{2005}]{Kulsrud2005}
{Kulsrud} R.~M.,  2005, {Plasma physics for astrophysics}

\bibitem[\protect\citeauthoryear{{Kulsrud} \& {Pearce}}{{Kulsrud} \&
  {Pearce}}{1969}]{Kulsrud1969}
{Kulsrud} R.,  {Pearce} W.~P.,  1969, \mn@doi [\apj] {10.1086/149981}, \href
  {http://adsabs.harvard.edu/abs/1969ApJ...156..445K} {156, 445}

\bibitem[\protect\citeauthoryear{{Lerche}}{{Lerche}}{1967}]{Lerche1967}
{Lerche} I.,  1967, \mn@doi [\apj] {10.1086/149045}, \href
  {http://adsabs.harvard.edu/abs/1967ApJ...147..689L} {147, 689}

\bibitem[\protect\citeauthoryear{{Navarro}, {Frenk}  \& {White}}{{Navarro}
  et~al.}{1996}]{NFW1}
{Navarro} J.~F.,  {Frenk} C.~S.,   {White} S.~D.~M.,  1996, \mn@doi [ApJ]
  {10.1086/177173}, \href {http://adsabs.harvard.edu/abs/1996ApJ...462..563N}
  {462, 563}

\bibitem[\protect\citeauthoryear{{Pakmor} \& {Springel}}{{Pakmor} \&
  {Springel}}{2013}]{Pakmor2013}
{Pakmor} R.,  {Springel} V.,  2013, \mn@doi [\mnras] {10.1093/mnras/stt428},
  \href {http://adsabs.harvard.edu/abs/2013MNRAS.432..176P} {432, 176}

\bibitem[\protect\citeauthoryear{{Pakmor}, {Bauer}  \& {Springel}}{{Pakmor}
  et~al.}{2011}]{Pakmor2011}
{Pakmor} R.,  {Bauer} A.,   {Springel} V.,  2011, \mn@doi [\mnras]
  {10.1111/j.1365-2966.2011.19591.x}, \href
  {http://adsabs.harvard.edu/abs/2011MNRAS.418.1392P} {418, 1392}

\bibitem[\protect\citeauthoryear{{Pakmor}, {Springel}, {Bauer}, {Mocz},
  {Munoz}, {Ohlmann}, {Schaal}  \& {Zhu}}{{Pakmor} et~al.}{2016a}]{Pakmor2016}
{Pakmor} R.,  {Springel} V.,  {Bauer} A.,  {Mocz} P.,  {Munoz} D.~J.,
  {Ohlmann} S.~T.,  {Schaal} K.,   {Zhu} C.,  2016a, \mn@doi [\mnras]
  {10.1093/mnras/stv2380}, \href
  {http://adsabs.harvard.edu/abs/2016MNRAS.455.1134P} {455, 1134}

\bibitem[\protect\citeauthoryear{{Pakmor}, {Pfrommer}, {Simpson}  \&
  {Springel}}{{Pakmor} et~al.}{2016b}]{Pakmor2016b}
{Pakmor} R.,  {Pfrommer} C.,  {Simpson} C.~M.,   {Springel} V.,  2016b, \mn@doi
  [\apjl] {10.3847/2041-8205/824/2/L30}, \href
  {http://adsabs.harvard.edu/abs/2016ApJ...824L..30P} {824, L30}

\bibitem[\protect\citeauthoryear{{Parrish} \& {Stone}}{{Parrish} \&
  {Stone}}{2005}]{Parrish2005}
{Parrish} I.~J.,  {Stone} J.~M.,  2005, \mn@doi [\apj] {10.1086/444589}, \href
  {http://adsabs.harvard.edu/abs/2005ApJ...633..334P} {633, 334}

\bibitem[\protect\citeauthoryear{{Pfrommer} \& {En{\ss}lin}}{{Pfrommer} \&
  {En{\ss}lin}}{2004}]{Pfrommer2004}
{Pfrommer} C.,  {En{\ss}lin} T.~A.,  2004, \mn@doi [\aap]
  {10.1051/0004-6361:20031464}, \href
  {http://adsabs.harvard.edu/abs/2004A%26A...413...17P} {413, 17}

\bibitem[\protect\citeauthoryear{{Pfrommer}, {Pakmor}, {Schaal}, {Simpson}  \&
  {Springel}}{{Pfrommer} et~al.}{2016}]{Pfrommer2016}
{Pfrommer} C.,  {Pakmor} R.,  {Schaal} K.,  {Simpson} C.~M.,   {Springel} V.,
  2016, preprint, \href {http://adsabs.harvard.edu/abs/2016arXiv160407399P} {}
  (\mn@eprint {arXiv} {1604.07399})

\bibitem[\protect\citeauthoryear{{Rodrigues}, {Sarson}, {Shukurov}, {Bushby}
  \& {Fletcher}}{{Rodrigues} et~al.}{2016}]{Rodrigues2016}
{Rodrigues} L.~F.~S.,  {Sarson} G.~R.,  {Shukurov} A.,  {Bushby} P.~J.,
  {Fletcher} A.,  2016, \mn@doi [\apj] {10.3847/0004-637X/816/1/2}, \href
  {http://adsabs.harvard.edu/abs/2016ApJ...816....2R} {816, 2}

\bibitem[\protect\citeauthoryear{{Ruszkowski}, {Yang}  \&
  {Zweibel}}{{Ruszkowski} et~al.}{2016}]{Ruszkowski2016}
{Ruszkowski} M.,  {Yang} H.-Y.~K.,   {Zweibel} E.,  2016, preprint, \href
  {http://adsabs.harvard.edu/abs/2016arXiv160204856R} {} (\mn@eprint {arXiv}
  {1602.04856})

\bibitem[\protect\citeauthoryear{Saad \& Schultz}{Saad \&
  Schultz}{1986}]{Saad1986}
Saad Y.,  Schultz M.~H.,  1986, \mn@doi [SIAM Journal on Scientific and
  Statistical Computing] {10.1137/0907058}, 7, 856

\bibitem[\protect\citeauthoryear{{Salem} \& {Bryan}}{{Salem} \&
  {Bryan}}{2014}]{Salem2014}
{Salem} M.,  {Bryan} G.~L.,  2014, \mn@doi [\mnras] {10.1093/mnras/stt2121},
  \href {http://adsabs.harvard.edu/abs/2014MNRAS.437.3312S} {437, 3312}

\bibitem[\protect\citeauthoryear{{Salem}, {Bryan}  \& {Hummels}}{{Salem}
  et~al.}{2014}]{Salem2014b}
{Salem} M.,  {Bryan} G.~L.,   {Hummels} C.,  2014, \mn@doi [\apjl]
  {10.1088/2041-8205/797/2/L18}, \href
  {http://adsabs.harvard.edu/abs/2014ApJ...797L..18S} {797, L18}

\bibitem[\protect\citeauthoryear{{Schlickeiser}}{{Schlickeiser}}{2002}]{Schlickeiser2002}
{Schlickeiser} R.,  2002, {Cosmic ray astrophysics}.
Springer., \url
  {http://esoads.eso.org/cgi-bin/nph-bib_query?bibcode=2002cra..book.....S&db_key=AST}

\bibitem[\protect\citeauthoryear{{Sharma} \& {Hammett}}{{Sharma} \&
  {Hammett}}{2007}]{Sharma2007}
{Sharma} P.,  {Hammett} G.~W.,  2007, \mn@doi [Journal of Computational
  Physics] {10.1016/j.jcp.2007.07.026}, \href
  {http://adsabs.harvard.edu/abs/2007JCoPh.227..123S} {227, 123}

\bibitem[\protect\citeauthoryear{{Sharma} \& {Hammett}}{{Sharma} \&
  {Hammett}}{2011}]{Sharma2011}
{Sharma} P.,  {Hammett} G.~W.,  2011, \mn@doi [Journal of Computational
  Physics] {10.1016/j.jcp.2011.03.009}, \href
  {http://adsabs.harvard.edu/abs/2011JCoPh.230.4899S} {230, 4899}

\bibitem[\protect\citeauthoryear{{Simpson}, {Pakmor}, {Marinacci}, {Pfrommer},
  {Springel}, {Glover}, {Clark}  \& {Smith}}{{Simpson}
  et~al.}{2016}]{Simpson2016}
{Simpson} C.~M.,  {Pakmor} R.,  {Marinacci} F.,  {Pfrommer} C.,  {Springel} V.,
   {Glover} S.~C.~O.,  {Clark} P.~C.,   {Smith} R.~J.,  2016, preprint, \href
  {http://adsabs.harvard.edu/abs/2016arXiv160602324S} {} (\mn@eprint {arXiv}
  {1606.02324})

\bibitem[\protect\citeauthoryear{{Skilling}}{{Skilling}}{1971}]{Skilling1971}
{Skilling} J.,  1971, \mn@doi [\apj] {10.1086/151210}, \href
  {http://adsabs.harvard.edu/abs/1971ApJ...170..265S} {170, 265}

\bibitem[\protect\citeauthoryear{{Sovinec} et~al.,}{{Sovinec}
  et~al.}{2004}]{Sovinec2004}
{Sovinec} C.~R.,  et~al., 2004, \mn@doi [Journal of Computational Physics]
  {10.1016/j.jcp.2003.10.004}, \href
  {http://adsabs.harvard.edu/abs/2004JCoPh.195..355S} {195, 355}

\bibitem[\protect\citeauthoryear{{Springel}}{{Springel}}{2010}]{Arepo}
{Springel} V.,  2010, \mn@doi [\mnras] {10.1111/j.1365-2966.2009.15715.x},
  \href {http://adsabs.harvard.edu/abs/2010MNRAS.401..791S} {401, 791}

\bibitem[\protect\citeauthoryear{{Springel} \& {Hernquist}}{{Springel} \&
  {Hernquist}}{2003}]{Springel2003}
{Springel} V.,  {Hernquist} L.,  2003, \mn@doi [\mnras]
  {10.1046/j.1365-8711.2003.06206.x}, \href
  {http://adsabs.harvard.edu/abs/2003MNRAS.339..289S} {339, 289}

\bibitem[\protect\citeauthoryear{{T{\"u}llmann}, {Dettmar}, {Soida}, {Urbanik}
  \& {Rossa}}{{T{\"u}llmann} et~al.}{2000}]{Tuellmann2000}
{T{\"u}llmann} R.,  {Dettmar} R.-J.,  {Soida} M.,  {Urbanik} M.,   {Rossa} J.,
  2000, \aap, \href {http://adsabs.harvard.edu/abs/2000A%26A...364L..36T} {364,
  L36}

\bibitem[\protect\citeauthoryear{{Uhlig}, {Pfrommer}, {Sharma}, {Nath},
  {En{\ss}lin}  \& {Springel}}{{Uhlig} et~al.}{2012}]{Uhlig2012}
{Uhlig} M.,  {Pfrommer} C.,  {Sharma} M.,  {Nath} B.~B.,  {En{\ss}lin} T.~A.,
  {Springel} V.,  2012, \mn@doi [\mnras] {10.1111/j.1365-2966.2012.21045.x},
  \href {http://adsabs.harvard.edu/abs/2012MNRAS.423.2374U} {423, 2374}

\bibitem[\protect\citeauthoryear{{Vogelsberger}, {Sijacki}, {Kere{\v s}},
  {Springel}  \& {Hernquist}}{{Vogelsberger} et~al.}{2012}]{Vogelsberger2012}
{Vogelsberger} M.,  {Sijacki} D.,  {Kere{\v s}} D.,  {Springel} V.,
  {Hernquist} L.,  2012, \mn@doi [\mnras] {10.1111/j.1365-2966.2012.21590.x},
  \href {http://adsabs.harvard.edu/abs/2012MNRAS.425.3024V} {425, 3024}

\bibitem[\protect\citeauthoryear{{Wiener}, {Oh}  \& {Guo}}{{Wiener}
  et~al.}{2013}]{Wiener2013}
{Wiener} J.,  {Oh} S.~P.,   {Guo} F.,  2013, \mn@doi [\mnras]
  {10.1093/mnras/stt1163}, \href
  {http://adsabs.harvard.edu/abs/2013MNRAS.434.2209W} {434, 2209}

\bibitem[\protect\citeauthoryear{{Yan} \& {Lazarian}}{{Yan} \&
  {Lazarian}}{2002}]{Yan2002}
{Yan} H.,  {Lazarian} A.,  2002, \mn@doi [Physical Review Letters]
  {10.1103/PhysRevLett.89.281102}, \href
  {http://adsabs.harvard.edu/abs/2002PhRvL..89B1102Y} {89, 1102}

\bibitem[\protect\citeauthoryear{{Yan} \& {Lazarian}}{{Yan} \&
  {Lazarian}}{2004}]{Yan2004}
{Yan} H.,  {Lazarian} A.,  2004, \mn@doi [\apj] {10.1086/423733}, \href
  {http://adsabs.harvard.edu/abs/2004ApJ...614..757Y} {614, 757}

\bibitem[\protect\citeauthoryear{van Leer}{van Leer}{1984}]{vanLeer1984}
van Leer B.,  1984, \mn@doi [SIAM Journal on Scientific and Statistical
  Computing] {10.1137/0905001}, 5, 1

\makeatother
\end{thebibliography}

\label{lastpage}

\end{document}